# Distributed Consensus Observers Based $H_\infty$ Control of Dissipative PDE Systems Using Sensor Networks

Huai-Ning Wu and Hong-Du Wang

***Abstract*—This paper considers the problem of finite dimensional output feedback $H_\infty$ control for a class of nonlinear spatially distributed processes described by highly dissipative partial differential equations (PDEs), whose state is observed by a sensor network (SN) with a given topology. This class of systems typically involves a spatial differential operator whose eigenspectrum can be partitioned into a finite-dimensional slow one and an infinite-dimensional stable fast complement. Motivated by this fact, the modal decomposition and singular perturbation techniques are initially applied to the PDE system to derive a finite dimensional ordinary differential equation model, which accurately captures the dominant dynamics of the PDE system. Subsequently, based on the slow system and the SN topology, a set of finite dimensional distributed consensus observers are constructed to estimate the state of the slow system. Then, a centralized control scheme, which only uses the available estimates from a specified group of SN nodes, is proposed for the PDE system. An $H_\infty$ control design is developed in terms of bilinear matrix inequality (BMI), such that the closed-loop PDE system is exponentially stable and a prescribed level of disturbance attenuation is satisfied for the slow system. Furthermore, a suboptimal $H_\infty$ controller is also provided to make the attenuation level as small as possible, which can be obtained via a local optimization algorithm that treats the BMI as double linear matrix inequality. Finally, the proposed method is applied to the control of one dimensional Kuramoto-Sivashinsky equation (KSE) system.**

## I. Introduction

THE past two decades have witnessed increasing focus on the analysis and control of spatially distributed processes (SDPs). A great deal of SDPs can be described by highly dissipative partial differential equations (PDEs), which contain the parabolic equation [1]-[3], the Kuramoto-Sivashinsky equation (KSE) [4], the Navier-Stokes equation (NSE) [5], to name a few. In general, physical phenomena that are described by dissipative PDEs include conduction during heat propagation, phased dynamics in reaction-diffusion systems and atmospheric pollution process over a given urban area.

Motivated by the fact that the dominant dynamic behavior of highly dissipative PDE systems can be characterized by a small number of degrees of freedom, most of the existing results on the control design for such systems involve initially the application of Galerkin's method to the PDE system to derive a low-dimensional ordinary differential equation (ODE) model, which is then used as the basis for the finite dimensional controller design purposes. For example, the finite-dimensional control problems of linear parabolic PDE systems were studied in [1], [6]-[8]. Recently, many nonlinear control methods have been also developed for dissipative PDE systems, including the geometric control [2], [4], [5], the fuzzy-model-based control [9], [10], adaptive neural control [11], and approximate optimal control [12], etc. In particular, some finite-dimensional control designs via dynamic output feedback (DOF) have been proposed for dissipative PDE systems [5]-[10]. However, it should be emphasized that, the existing DOF control results in [5]-[10] were developed on the basis of a finite-dimensional centralized observer. To the best of the authors' knowledge, very little attention has been paid to the finite-dimensional distributed observers based control design for nonlinear dissipative PDE systems.

On the other hand, significant advances in science and technology lead to a large number of SDPs that are often deployed in large and involve numerous sensors, actuators and internal process variables [13]-[15]. In practice, the set of sensor nodes with computation and communication capacity distributed along the spatial extent of the process usually form a sensor network (SN). Recently, many engineering applications have considered the use of SNs to provide efficient and effective remote monitoring/control of SDPs. Examples include the monitoring and elimination of diffusion pollutions using mobile SNs [15] and the structural health monitoring of buildings and bridges [16]. However, in the case of using SNs, a centralized observer may be impractical or impossible for the state estimation or control, due to high dimensionality of the target system or the limit of power supply and communication capacity of the sensor nodes. In order to overcome this difficulty, the consensus-based distributed estimation problem of SNs has gained rapidly increasing interest in the past few years (see, e.g., [17]-[21], and the references therein), whose objective is to develop a set of distributed local observers for achieving a

Mauscript received June 31, 2013. This work was supported in part by the National Basic Research Program of China (973 Program) under Grant 2012CB720003 and in part by the National Natural Science Foundations of China under Grants 61121003 and 61473011.

The authors are with the Science and Technology on Aircraft Control Laboratory, School of Automation Science and Electrical Engineering, Beihang University, Beijing 100191, China (e-mail: whn@buaa.edu.cn)

common estimate of state at each SN node. These observers are usually called distributed consensus observers (DCOs). Compared with the centralized estimation approach, the distributed one has its own advantages such as low communication burden, fast implementation, and low cost [17]. Until now, the existing works on distributed consensus estimation have been mainly developed for ODE systems, most of which focused on proposing different mechanisms for combining the Kalman filter [18] or $H_\infty$ filter [19] with a consensus filter to enforce the consensus of the estimation outcomes of all local filters. As regards SDPs, more recently, some distributed consensus estimation schemes have been proposed in [20] and [21], which enforce consensus of the spatially distributed estimators by dynamically minimizing the disagreement between them. Despite these efforts, however, very little research has directly addressed the problem of $H_\infty$ control design based on DCOs for a nonlinear SDP by using an SN with given topology, which motivates this study.

In this paper, we are concerned with the problem of finite dimensional DCOs-based $H_\infty$ control design for nonlinear dissipative PDE systems with SNs. The modal decomposition and singular perturbation techniques are initially applied to the PDE system to derive a finite dimensional ODE model, which accurately captures the dynamics of the dominant (slow) modes of the PDE system. Subsequently, based on the slow system and the SN topology, a set of finite dimensional DCOs are constructed to estimate the state of the slow system and enforce the agreement of all estimates. Then, an $H_\infty$ control design method is developed in terms of bilinear matrix inequality (BMI) to ensure the closed-loop exponential stability of the original PDE system while satisfying a prescribed level of disturbance attenuation for the slow system. Furthermore, to make the attenuation level as small as possible, a suboptimal $H_\infty$ controller design problem is also addressed, which can be solved by a local optimization algorithm that treats the BMI as double linear matrix inequality (LMI). Finally, a simulation study on the control of one dimensional KSE system is given to show the effectiveness of the proposed design method.

The main innovations and contributions can be summarized as follows. 1) This paper aims at solving the DCOs-based $H_\infty$ control design problem for a class of nonlinear dissipative PDE systems via an SN. To the best authors' knowledge, this problem is rarely studied. 2) A set of finite dimensional DCOs accounting for the complex communication between sensor nodes are proposed to compute the slow mode estimates for the control design. 3) Compared with the existing infinite dimensional results in [20] and [21], the developed finite dimensional control result can significantly improve the computational efficiency and reduce the communication burden of the SNs.

*Notations*: $\mathbb{R}$ and $\mathbb{R}_+$ denote the set of real and nonnegative real numbers, respectively. $\mathbb{R}^n$ and $\mathbb{R}^{n\times m}$ are the $n$-dimensional Euclidean space and the set of all real $n\times m$ matrices, respectively. $|\cdot|$ and $\|\cdot\|$ stand for the absolute value for scalars and Euclidean norm for vectors, respectively. The superscript $T$ is used for the transpose. Let $l^2$ denote the subset of $\mathbb{R}^\infty$ consisting of all square summable infinite sequences of real numbers, so that $l^2=\{\boldsymbol{x}=[x_1\ \cdots\ x_\infty]^T\in\mathbb{R}^\infty:\|\boldsymbol{x}\|_{l^2}<\infty\}$ where $\|\boldsymbol{x}\|_{l^2}\triangleq\sqrt{\sum_{i=1}^{\infty}x_i^2}$ . For a symmetric matrix $\boldsymbol{M}$ , $\boldsymbol{M}>(\geq,<,\leq)0$ means that it is positive definite (positive semi-definite, negative definite, negative semi-definite, respectively). $\lambda_{\min}(\cdot)$ ( $\lambda_{\max}(\cdot)$ ) denotes the minimum (maximum) eigenvalue of a matrix. The identity matrix of dimension $n$ is denoted by $\boldsymbol{I}_n$ (or $\boldsymbol{I}$, if the dimension is clear from the context). The $N$-dimensional column vector of 1's is denoted by $1_N$ . $\boldsymbol{A}\otimes\boldsymbol{B}\in\mathbb{R}^{mp\times nq}$ is the Kronecker product of matrices $\boldsymbol{A}\in\mathbb{R}^{m\times n}$ and $\boldsymbol{B}\in\mathbb{R}^{p\times q}$ . $\mathrm{diag}_{i=1}^n\{\boldsymbol{A}_i\}$ and $\mathrm{col}_{i=1}^n\{\boldsymbol{A}_i\}$ denote the block diagonal matrix $\mathrm{diag}\{\boldsymbol{A}_1,\ldots,\boldsymbol{A}_n\}$ and the block column vector of $n$ block matrices $\boldsymbol{A}_i$ , $i=1,\cdots,m$ , respectively. The symbol $*$ is used as an ellipsis for terms in matrix expressions that are induced by symmetry, e.g.,

$$\begin{bmatrix}\boldsymbol{S}+[\boldsymbol{M}+*] & * \\ \boldsymbol{X} & \boldsymbol{Y}\end{bmatrix}\triangleq\begin{bmatrix}\boldsymbol{S}+[\boldsymbol{M}+\boldsymbol{M}^T] & \boldsymbol{X} \\ \boldsymbol{X}^T & \boldsymbol{Y}\end{bmatrix}.$$

## II. Preliminaries and Problem Statement

### A. Description of dissipative PDE Systems with SNs

We consider a class of SDPs described by the following highly dissipative and nonlinear PDEs:

$$\frac{\partial\bar{x}(z,t)}{\partial t}=\mathcal{A}\bar{x}(z,t)+f(\bar{x}(z,t))+k_u\boldsymbol{b}_u^T(z)\boldsymbol{u}(t)+k_w\boldsymbol{b}_w^T(z)\boldsymbol{w}(t) \quad (1)$$

subject to the boundary conditions

$$\boldsymbol{l}(t,\bar{x},\frac{\partial\bar{x}}{\partial z},\cdots,\frac{\partial^{n_z-1}\bar{x}}{\partial z^{n_z-1}})=0 \ \text{ on } \Gamma \quad (2)$$

and the initial condition

$$\bar{x}(z,t)=\bar{x}_0(z) \quad (3)$$

where $\bar{x}(z,t)\in\mathbb{R}$ is the vector of state variables, $t\geq 0$ is the time variable, $z\in\Omega\triangleq[z_1,z_2]\subset\mathbb{R}$ is the spatial variable, $\Omega$ is the spatial domain of definition of the SDP and $\Gamma$ is its boundary, $\boldsymbol{u}(t)\in\mathbb{R}^{q_u}$ is the manipulated input vector of the actuators, and $\boldsymbol{w}(t)\in\mathbb{R}^{q_w}$ denotes the bounded process disturbance. $\mathcal{A}$ is a dissipative, self-adjoint, linear spatial differential operator of the form

$$\mathcal{A}=a_1\frac{\partial}{\partial z}+a_2\frac{\partial^2}{\partial z^2}+\cdots+a_{n_z}\frac{\partial^{n_z}}{\partial z^{n_z}}$$

in which $a_i$ , $i=1,2,\cdots,n_z$ are known constants , $n_z$ is the highest order of spatial derivatives in the PDE and usually an even number (e.g., $n_z=2$ for the parabolic PDE [2] and the NSE [5], $n_z=4$ for the KSE [4]). $f$ is a locally Lipschitz continuous nonlinear function satisfying $f(0)=0$ . $k_u$ and $k_w$ are known constants. $\boldsymbol{b}_u(z)=[b_{u,1}(z)\ \cdots\ b_{u,q_u}(z)]^T$ and $\boldsymbol{b}_w(z)=[b_{w,1}(z)\ \cdots\ b_{w,q_w}(z)]^T$ are known smooth vector functions of $z$, where $b_{u,i}(z)$ denotes how the control action $u_i(t)$ is distributed in $\Omega$, $b_{w,i}(z)$ specifies the position of action of the exogenous disturbance $w_i(t)$ in $\Omega$. $\boldsymbol{l}$ is a sufficiently smooth nonlinear vector function, $\partial\bar{x}/\partial z|_\Gamma$ is the normal derivative on the boundary $\Gamma$, and $\bar{x}_0(z)$ is a smooth vector function of $z$.

The state of the SDP (1) is observed by an SN of $p$ nodes distributed along the spatial extent of the SDP, whose sensing models are given by

$$\boldsymbol{y}_i(t)=\int_\Omega \boldsymbol{s}_i(z)\bar{x}(z,t)dz+\boldsymbol{v}_i(t),\ i\in\mathcal{S}\triangleq\{1,2,\cdots,p\} \tag{4}$$

where $\boldsymbol{y}_i(t)\in\mathbb{R}^{q_{y,i}}$ is the measured output of the $i$-th node equipped with $q_{y,i}$ sensors, $\boldsymbol{v}_i(t)\in\mathbb{R}^{q_{y,i}}$ is the bounded measurement disturbance, and $\boldsymbol{s}_i(z)=[s_{i1}(z),\cdots,s_{iq_{y,i}}(z)]^T$ is a known smooth vector function of $z$, where $s_{ij}(z)$ is determined by the location and shape (point or distributed) of the $j$-th sensor in the $i$-th node.

The topology of the SN can be represented by a direct graph $\mathcal{G}=(\mathcal{S},\mathcal{E},\mathcal{M})$ of order $p$ with the set of nodes $\mathcal{S}$, the set of edges $\mathcal{E}\subseteq\mathcal{S}\times\mathcal{S}$, and the weighted adjacency matrix $\mathcal{M}=[m_{ij}]_{p\times p}$. An edge of $\mathcal{G}$ is denoted by $(i,j)$. The adjacent elements associated with the edges of the graph are positive, i.e., $m_{ij}>0\Leftrightarrow(i,j)\in\mathcal{E}$. Moreover, we assume $m_{ii}=0$ for all $i\in\mathcal{S}$. The set of neighbors of the $i$-th SN node is denoted by $\mathcal{N}_i=\{j\in\mathcal{S}:(i,j)\in\mathcal{E}\}$.

### *B. Infinite-dimensional singular perturbation formulation of the PDE system*

To simplify the presentation, we define the Hilbert space $\mathcal{H}_{2,\Omega}\triangleq\left\{\phi:\Omega\mapsto\mathbb{R}\text{ and }\|\phi\|_{2,\Omega}<\infty\right\}$ with inner product $\langle\phi_1,\phi_2\rangle\triangleq\int_\Omega\phi_1(z)\phi_2(z)dz$ and norm $\|\phi_1\|_{2,\Omega}\triangleq\langle\phi_1,\phi_1\rangle^{\frac{1}{2}}$, where $\phi_1$ and $\phi_2$ are two elements of $\mathcal{H}_{2,\Omega}$. The domain of the operator $\mathcal{A}$ is denoted by

$$\mathcal{D}(\mathcal{A})\triangleq\{\ \phi\in\mathcal{H}_{2,\Omega}\text{ and }\boldsymbol{l}(t,\phi,\frac{\partial\phi}{\partial z},\cdots,\frac{\partial^{n_z-1}\phi}{\partial z^{n_z-1}})=0\text{ on }\Gamma\ \}.$$

To present the theoretical results, the PDE system of (1)-(3) will be formulated as an infinite dimensional singular perturbation model of ODEs through modal decomposition technique. For the operator $\mathcal{A}$, the eigenvalue problem is defined as $\mathcal{A}\phi_j(z)=\lambda_j\phi_j(z)$, $j=1,2,\cdots,\infty$ where $\lambda_j$ is the $j$-th eigenvalue and $\phi_j(z)\in\mathcal{D}(\mathcal{A})$ is the corresponding orthonormal eigenfunction, i.e., $\langle\phi_k(z),\phi_j(z)\rangle=\delta(k-j)$, in which $\delta(\cdot)$ is the Kronecker delta function. These eigenfunctions form an orthonormal basis for domain $\mathcal{D}(\mathcal{A})$. Moreover, all eigenvalues of the self-ajoint operator $\mathcal{A}$ are real. To facilitate the subsequent development, we give the following assumption.

*Assumption 1*: All eigenvalues of $\mathcal{A}$ are ordered so that $\lambda_j\geq\lambda_{j+1}$, and there is a finite number $n$ so that $\lambda_{n+1}<0$ and $\varepsilon\triangleq|\lambda_L|/|\lambda_{n+1}|<1$ is a small positive number, where $\lambda_L$ is the largest non-zero eigenvalue.

Expand the solution of the system of (1) into an infinite series in terms of the basis functions $\phi_j(z)$ as follows:

$$\bar{x}(z,t)=\sum_{j=1}^{\infty}x_j(t)\phi_j(z)=\boldsymbol{\phi}_s^T(z)\boldsymbol{x}_s(t)+\boldsymbol{\phi}_f^T(z)\boldsymbol{x}_f(t) \tag{5}$$

where $x_j(t)$ ( $j=1,2,\cdots,\infty$ ) are time-varying coefficients called the modes of the PDE system, $\boldsymbol{\phi}_s(z)\triangleq[\phi_1(z)\ \cdots\ \phi_n(z)]^T$, $\boldsymbol{\phi}_f(z)\triangleq[\phi_{n+1}(z)\ \cdots\ \phi_\infty(z)]^T$, $\boldsymbol{x}_s(t)\triangleq[x_1(t)\ \cdots\ x_n(t)]^T\in\mathbb{R}^n$ and $\boldsymbol{x}_f(t)\triangleq[x_{n+1}(t)\ \cdots\ x_\infty(t)]^T\in l^2$ are the slow and fast mode vectors, respectively. Taking the inner product of both sides of (5) with $\phi_j(z)$, we can immediately write down the following relation:

$$x_j(t)=\langle\bar{x}(\cdot,t),\phi_j(\cdot)\rangle. \tag{6}$$

Differentiating both sides of (6) with respect to time and considering (1), (5) and (6) give

$$\frac{dx_j(t)}{dt}=\lambda_j x_j(t)+\langle f(\bar{x}(\cdot,t)),\phi_j(\cdot)\rangle+\boldsymbol{b}_{u,j}^T\boldsymbol{u}(t)+\boldsymbol{b}_{w,j}^T\boldsymbol{w}(t),$$
$$x_j(0)=x_{j,0},\ j=1,2,\cdots,\infty$$

where $x_{j,0}=\langle\bar{x}_0(\cdot),\phi_j(\cdot)\rangle$ and

$$\boldsymbol{b}_{u,j}=\left[\langle k_u b_{u,1}(z),\phi_j(\cdot)\rangle\ \cdots\ \langle k_u b_{u,q_u}(\cdot),\phi_j(\cdot)\rangle\right]^T,$$
$$\boldsymbol{b}_{w,j}=\left[\langle k_w b_{w,1}(z),\phi_j(\cdot)\rangle\ \cdots\ \langle k_w b_{w,q_u}(\cdot),\phi_j(\cdot)\rangle\right]^T,$$

which can be rewritten as the following infinite dimensional ODE system consisting of the slow and fast subsystems:

$$\begin{cases}\dot{\boldsymbol{x}}_s=\boldsymbol{A}_s\boldsymbol{x}_s+\boldsymbol{f}_s(\boldsymbol{x}_s,\boldsymbol{x}_f)+\boldsymbol{B}_{u,s}\boldsymbol{u}+\boldsymbol{B}_{w,s}\boldsymbol{w},\ \boldsymbol{x}_s(0)=\boldsymbol{x}_{s,0}\\ \dot{\boldsymbol{x}}_f=\boldsymbol{A}_f\boldsymbol{x}_f+\boldsymbol{f}_f(\boldsymbol{x}_s,\boldsymbol{x}_f)+\boldsymbol{B}_{u,f}\boldsymbol{u}+\boldsymbol{B}_{w,f}\boldsymbol{w},\ \boldsymbol{x}_f(0)=\boldsymbol{x}_{f,0}\end{cases} \tag{7}$$

where

$$\boldsymbol{A}_s=\text{diag}\{\lambda_1,\cdots,\lambda_n\},\ \boldsymbol{A}_f=\text{diag}\{\lambda_{n+1},\cdots,\lambda_\infty\},$$

$$\boldsymbol{f}_s(\boldsymbol{x}_s,\boldsymbol{x}_f)=\begin{bmatrix}f_1(\boldsymbol{x}_s,\boldsymbol{x}_f)\\ \vdots\\ f_n(\boldsymbol{x}_s,\boldsymbol{x}_f)\end{bmatrix},\ \boldsymbol{f}_f(\boldsymbol{x}_s,\boldsymbol{x}_f)=\begin{bmatrix}f_{n+1}(\boldsymbol{x}_s,\boldsymbol{x}_f)\\ \vdots\\ f_\infty(\boldsymbol{x}_s,\boldsymbol{x}_f)\end{bmatrix},$$

$$\boldsymbol{B}_{u,s}=\begin{bmatrix}\boldsymbol{b}_{u,1}^T\\ \vdots\\ \boldsymbol{b}_{u,n}^T\end{bmatrix},\ \boldsymbol{B}_{u,f}=\begin{bmatrix}\boldsymbol{b}_{u,n+1}^T\\ \vdots\\ \boldsymbol{b}_{u,\infty}^T\end{bmatrix},\ \boldsymbol{B}_{w,s}=\begin{bmatrix}\boldsymbol{b}_{w,1}^T\\ \vdots\\ \boldsymbol{b}_{w,n}^T\end{bmatrix},\ \boldsymbol{B}_{w,f}=\begin{bmatrix}\boldsymbol{b}_{w,n+1}^T\\ \vdots\\ \boldsymbol{b}_{w,\infty}^T\end{bmatrix},$$

$$\boldsymbol{x}_{s,0}=[x_{1,0}\ \cdots\ x_{n,0}]^T,\ \boldsymbol{x}_{f,0}=[x_{n+1,0}\ \cdots\ x_{\infty,0}]^T$$

with $f_j(\boldsymbol{x}_s,\boldsymbol{x}_f)=\langle f(\boldsymbol{\phi}_s^T(\cdot)\boldsymbol{x}_s+\boldsymbol{\phi}_f^T(\cdot)\boldsymbol{x}_f),\phi_j(\cdot)\rangle$.

Then, multiplying the fast subsystem by a small positive parameter $\varepsilon$ yields the following singular perturbation model of the dynamical system (7):

$$\begin{cases} \dot{\boldsymbol{x}}_s = \boldsymbol{A}_s \boldsymbol{x}_s + \boldsymbol{f}_s(\boldsymbol{x}_s, \boldsymbol{x}_f) + \boldsymbol{B}_{u,s}\boldsymbol{u} + \boldsymbol{B}_{w,s}\boldsymbol{w} \\ \varepsilon \dot{\boldsymbol{x}}_f = \boldsymbol{A}_{f\varepsilon}\boldsymbol{x}_f + \varepsilon \boldsymbol{f}_f(\boldsymbol{x}_s, \boldsymbol{x}_f) + \varepsilon \boldsymbol{B}_{u,f}\boldsymbol{u} + \varepsilon \boldsymbol{B}_{w,f}\boldsymbol{w} \end{cases} \tag{8}$$

where $\boldsymbol{A}_{f\varepsilon} = \varepsilon \boldsymbol{A}_f$ .

As a consequence, the singular perturbation theory [22] can be applied for our study. By introducing the fast time-scale $\zeta = t/\varepsilon$ and setting $\varepsilon = 0$, the following infinite dimensional fast subsystem is obtained from the model (8):

$$\mathrm{d}\boldsymbol{x}_f(\zeta)/d\zeta = \boldsymbol{A}_{f\varepsilon}\boldsymbol{x}_f(\zeta). \tag{9}$$

It follows from the fact $\lambda_{n+1} < 0$ and the definition of $\varepsilon$ that the fast system (9) is globally exponentially stable. Setting $\varepsilon = 0$ in (8), we get the unique root $\boldsymbol{x}_f = 0$ due to the nonsingularity of $\boldsymbol{A}_{f\varepsilon}$. Substituting $\boldsymbol{x}_f = 0$ into (8) yields the following finite dimensional slow subsystem:

$$\dot{\boldsymbol{x}}_s = \boldsymbol{A}_s \boldsymbol{x}_s + \boldsymbol{f}_s(\boldsymbol{x}_s, 0) + \boldsymbol{B}_{u,s}\boldsymbol{u} + \boldsymbol{B}_{w,s}\boldsymbol{w}. \tag{10}$$

Using (5), the measurement equations in (4) are given as

$$\boldsymbol{y}_i = \boldsymbol{C}_{s,i}\boldsymbol{x}_s + \boldsymbol{C}_{f,i}\boldsymbol{x}_f + \boldsymbol{v}_i = \boldsymbol{C}_{s,i}\boldsymbol{x}_s + \bar{\boldsymbol{v}}_i, \; i \in \mathcal{S} \tag{11}$$

where $\boldsymbol{C}_{s,i} = \int_\Omega \boldsymbol{s}_i(z)\boldsymbol{\phi}_s^T(z)dz$ , $\boldsymbol{C}_{f,i} = \int_\Omega \boldsymbol{s}_i(z)\boldsymbol{\phi}_f^T(z)dz$ , and $\bar{\boldsymbol{v}}_i \triangleq \boldsymbol{C}_{f,i}\boldsymbol{x}_f + \boldsymbol{v}_i$ is the measurement disturbance of the slow system in the $i$-th node. $\boldsymbol{y}_{f,i} \triangleq \boldsymbol{C}_{f,i}\boldsymbol{x}_f$ is usually referred to as the observation spillover. The slow system (10) with measurement equations in (11) will be used as the basis for the control design for PDE system (1)-(4) in this study.

### C. Problem statement

Assume that the pairs $(\boldsymbol{A}_s, \boldsymbol{C}_{s,i})$, $i \in \mathcal{S}$ are observable, i.e., the PDE system is approximately observable [23] for each node $i$ , $i \in \mathcal{S}$ . Then, based on the slow system (10) and the measurement equations in (11), we consider the following $p$ finite dimensional local Luenberger-like DCOs:

$$\begin{aligned} \dot{\hat{\boldsymbol{x}}}_{s,i}(t) = {} & \boldsymbol{A}_s\hat{\boldsymbol{x}}_{s,i}(t) + \boldsymbol{B}_{u,s}\boldsymbol{u}(t) + \boldsymbol{L}_i(\boldsymbol{y}_i(t) - \boldsymbol{C}_{s,i}\hat{\boldsymbol{x}}_{s,i}(t)) \\ & + \sum_{j\in\mathcal{N}_i} m_{ij}\boldsymbol{G}_{ij}(\hat{\boldsymbol{x}}_{s,i}(t) - \hat{\boldsymbol{x}}_{s,j}(t)), \; \hat{\boldsymbol{x}}_{s,i}(0) = 0, \; i \in \mathcal{S} \end{aligned} \tag{12}$$

where $\hat{\boldsymbol{x}}_{s,i} = [\hat{x}_{1,i} \; \cdots \; \hat{x}_{n,i}]^T \in \mathbb{R}^n$ is the estimate of $\boldsymbol{x}_s$ provided by the local observer in the $i$-th node, $\boldsymbol{L}_i \in \mathbb{R}^{n\times q_{y,i}}$ and $\boldsymbol{G}_{ij} \in \mathbb{R}^{n\times n}$ for $j \in \mathcal{S}$, $j \neq i$, are the Luenberger and consensus gain matrices of the local observer, respectively.

*Remark 1:* The local distributed observers in (12) account for the communications between the underlying node and its neighboring nodes. Once the Luenberger and consensus gain matrices of all observers are determined, the state estimation algorithm for the slow system can be computed via the SN in a distributed manner.

*Remark 2*: It is worth mentioning that the distributed consensus estimation methods of linear SDPs in [20] and [21] are developed in an abstract framework. These infinite dimensional methods may lead to a major challenge for numerical implementation and computation complexity with the high dimensionality of the approximation of the underlying SDP. In this study, a set of computationally efficient finite dimensional DCOs are constructed to estimate the slow modes of dissipative PDE systems, which can reduce the communication burden of the SN significantly.

We consider the following DCOs-based centralized controller:

$$\boldsymbol{u}(t) = \sum_{i\in\mathcal{U}} \boldsymbol{K}_i \hat{\boldsymbol{x}}_{s,i}(t) \tag{13}$$

where $\boldsymbol{K}_i \in \mathbb{R}^{q_u\times n}$ , $i \in \mathcal{U} \subseteq \mathcal{S}$ are control gain matrices to be determined, $\mathcal{U}$ is a subset of $\mathcal{S}$ representing the set of the nodes that can transmit the estimates to the controller for computing the control inputs of the actuators. Fig. 1 shows the diagram of the DCOs-based centralized controller for the SDP with an SN.

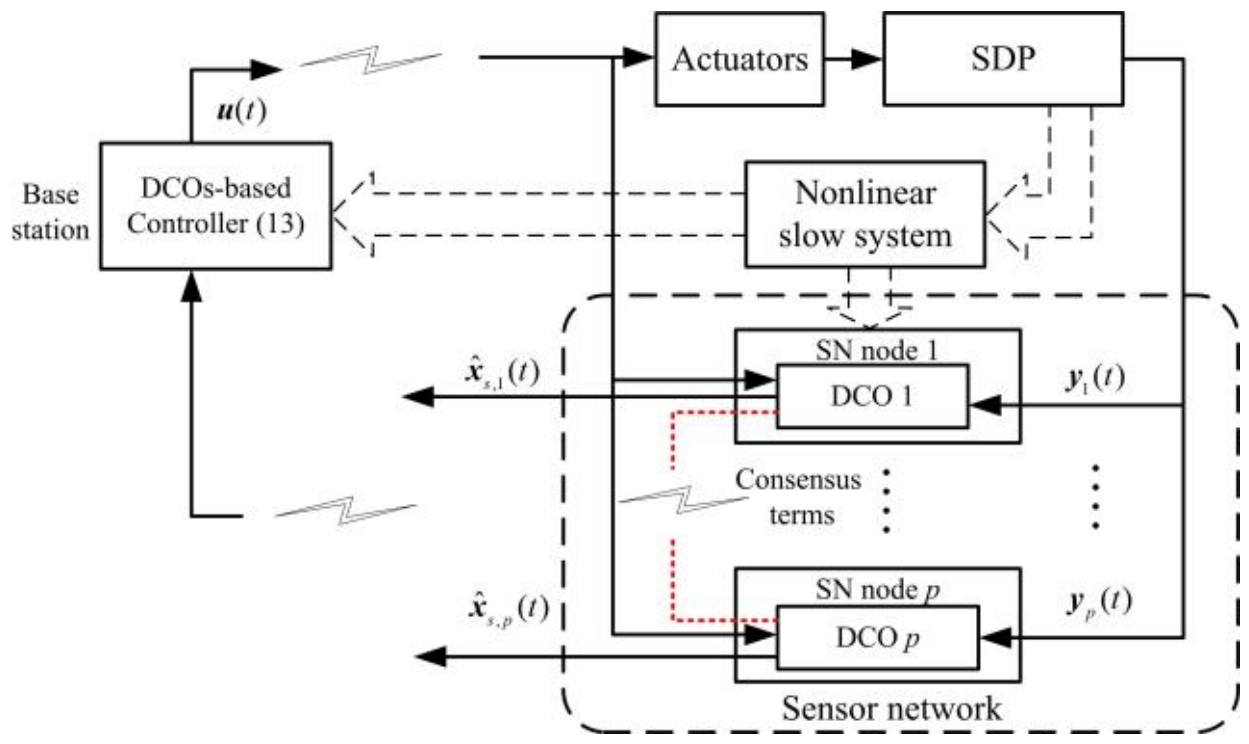

Fig. 1 Finite dimensional DCOs-based centralized controller

As is well known, $H_\infty$ control is an effective control methodology to attenuate the effect of uncertain external disturbance on the desired control performance. Thus, here we consider the following $H_\infty$ control performance index for the slow system (10) under zero-initial condition (i.e., $\bar{\boldsymbol{x}}_0(z) = 0$ ):

$$\int_0^{t_f} [\boldsymbol{x}_s^T(t)\boldsymbol{Q}\boldsymbol{x}_s(t) + \boldsymbol{u}^T(t)\boldsymbol{R}\boldsymbol{u}(t)]dt \le \gamma^2 \int_0^{t_f} \tilde{\boldsymbol{w}}^T(t)\tilde{\boldsymbol{w}}(t)dt \tag{14}$$

where $\tilde{\boldsymbol{w}} = [\boldsymbol{w}^T \quad \bar{\boldsymbol{v}}^T]^T$ and $\bar{\boldsymbol{v}} = \mathrm{col}_{i=1}^p\{\bar{\boldsymbol{v}}_i\}$ , $t_f$ is the final time of control, $\boldsymbol{Q} \ge 0$ , $\boldsymbol{R} = \boldsymbol{D}_R^T\boldsymbol{D}_R \ge 0$ are given weighting matrices, and $\gamma > 0$ is a prescribed attenuation level.

*Remark 3*: It should be pointed out that the performance (14) can be transformed into an $H_\infty$ performance for the original PDE system by making some additional assumptions in a similar way as in [24].

Therefore, the problem under consideration is to find a centralized controller of the form (13) based on the finite dimensional DCOs in (12), such that the closed-loop PDE system is exponentially stable in the absence of disturbances $\boldsymbol{w}$ and $\boldsymbol{v}_i$, $i \in \mathcal{S}$, and the $H_\infty$ control performance in (14) is achieved in the presence of disturbance $\tilde{\boldsymbol{w}}$. In general, it is desirable to make the attenuation level as small as possible.

To facilitate this study, we make the following assumption.

*Assumption 2:* There exists a known positive constant $\kappa_1$ such that the nonlinear function $\boldsymbol{f}_s(\boldsymbol{x}_s, 0)$ in (10) satisfies $\|\boldsymbol{f}_s(\boldsymbol{x}_s, 0)\| \le \kappa_1 \|\boldsymbol{x}_s\|$.

## III. Finite Dimensional DCOS-Based $H_\infty$ Control Design

For convenience, we let

$$\bar{\boldsymbol{G}}_{ij} = m_{ij}\boldsymbol{G}_{ij},\ i, j \in \mathcal{S}. \tag{15}$$

Then, (12) can be rewritten as

$$\dot{\hat{\boldsymbol{x}}}_{s,i} = \boldsymbol{A}_s\hat{\boldsymbol{x}}_{s,i} + \boldsymbol{B}_{u,s}\boldsymbol{u} + \boldsymbol{L}_i\boldsymbol{C}_{s,i}(\boldsymbol{x}_s - \hat{\boldsymbol{x}}_{s,i}) + \sum_{j\in\mathcal{S}}\bar{\boldsymbol{G}}_{ij}(\hat{\boldsymbol{x}}_{s,i} - \hat{\boldsymbol{x}}_{s,j}) + \boldsymbol{L}_i\bar{\boldsymbol{v}}_i,\ \hat{\boldsymbol{x}}_{s,i}(0) = 0,\ i \in \mathcal{S}. \tag{16}$$

Setting $\boldsymbol{K}_i \equiv 0$ when $i \notin \mathcal{U}$, we can write (13) as

$$\boldsymbol{u} = \sum_{i\in\mathcal{U}}\boldsymbol{K}_i\hat{\boldsymbol{x}}_{s,i} = \sum_{i\in\mathcal{S}}\boldsymbol{K}_i\hat{\boldsymbol{x}}_{s,i}. \tag{17}$$

Denoting $\boldsymbol{e}_{s,i} = \boldsymbol{x}_s - \hat{\boldsymbol{x}}_{s,i}$ and subtracting (16) from (10) give

$$\dot{\boldsymbol{e}}_{s,i} = (\boldsymbol{A}_s - \boldsymbol{L}_i\boldsymbol{C}_{s,i} + \sum_{j\in\mathcal{S}}\bar{\boldsymbol{G}}_{ij})\boldsymbol{e}_{s,i} - \sum_{j\in\mathcal{S}}\bar{\boldsymbol{G}}_{ij}\boldsymbol{e}_{s,j} + \boldsymbol{f}_s(\boldsymbol{x}_s, 0) + \boldsymbol{B}_{w,s}\boldsymbol{w}(t) - \boldsymbol{L}_i\bar{\boldsymbol{v}}_i,\ \boldsymbol{e}_{s,i}(0) = \boldsymbol{x}_{s,0}. \tag{18}$$

Defining $\bar{\boldsymbol{e}}_s \triangleq [\boldsymbol{e}_{s,1}^T \cdots \boldsymbol{e}_{s,p}^T]^T \in \mathbb{R}^{np}$ and using (18), we have

$$\dot{\bar{\boldsymbol{e}}}_s = (\bar{\boldsymbol{A}}_s - \bar{\boldsymbol{L}}\bar{\boldsymbol{C}}_s + \bar{\boldsymbol{G}})\bar{\boldsymbol{e}}_s + (1_p \otimes \boldsymbol{I}_n)\boldsymbol{f}_s(\boldsymbol{x}_s, 0) + (1_p \otimes \boldsymbol{B}_{w,s})\boldsymbol{w} - \bar{\boldsymbol{L}}\bar{\boldsymbol{v}} \tag{19}$$

where $\bar{\boldsymbol{A}}_s = \boldsymbol{I}_n \otimes \boldsymbol{A}_s$, $\bar{\boldsymbol{L}} = \text{diag}_{i=1}^p\{\boldsymbol{L}_i\}$, $\bar{\boldsymbol{C}}_s = \text{diag}_{i=1}^p\{\boldsymbol{C}_{s,i}\}$, and

$$\bar{\boldsymbol{G}} = \begin{bmatrix} \sum_{j\in\mathcal{S}}\bar{\boldsymbol{G}}_{1j} & -\bar{\boldsymbol{G}}_{12} & \cdots & -\bar{\boldsymbol{G}}_{1p} \\ -\bar{\boldsymbol{G}}_{21} & \sum_{j\in\mathcal{S}}\bar{\boldsymbol{G}}_{2j} & \cdots & -\bar{\boldsymbol{G}}_{2p} \\ \vdots & \vdots & \ddots & \vdots \\ -\bar{\boldsymbol{G}}_{p1} & -\bar{\boldsymbol{G}}_{p2} & \cdots & \sum_{j\in\mathcal{S}}\bar{\boldsymbol{G}}_{pj} \end{bmatrix},\ \bar{\boldsymbol{x}}_{s,0} = 1_p \otimes \boldsymbol{x}_{s,0}.$$

From (10), (17) and (19), we can obtain the following augmented closed-loop system:

$$\dot{\tilde{\boldsymbol{x}}}_s = \tilde{\boldsymbol{A}}\tilde{\boldsymbol{x}}_s + (1_{p+1} \otimes \boldsymbol{I}_n)\boldsymbol{f}_s(\boldsymbol{x}_s, 0) + \tilde{\boldsymbol{B}}\tilde{\boldsymbol{w}} \tag{20}$$

where

$$\tilde{\boldsymbol{x}}_s = \begin{bmatrix} \boldsymbol{x}_s \\ \bar{\boldsymbol{e}}_s \end{bmatrix} \in \mathbb{R}^{n(p+1)},\ \tilde{\boldsymbol{A}} = \begin{bmatrix} \boldsymbol{A}_s + \boldsymbol{B}_{u,s}\sum_{i\in\mathcal{S}}\boldsymbol{K}_i & -\boldsymbol{B}_{u,s}\bar{\boldsymbol{K}} \\ 0 & \bar{\boldsymbol{A}}_s - \bar{\boldsymbol{L}}\bar{\boldsymbol{C}}_s + \bar{\boldsymbol{G}} \end{bmatrix},$$

$$\tilde{\boldsymbol{B}} = \begin{bmatrix} \boldsymbol{B}_{w,s} & 0 \\ 1_p \otimes \boldsymbol{B}_{w,s} & -\bar{\boldsymbol{L}} \end{bmatrix},\ \bar{\boldsymbol{K}} = [\boldsymbol{K}_1 \ \cdots \ \boldsymbol{K}_p].$$

It is observed that $\bar{\boldsymbol{G}}_{ij} = 0$ when $m_{ij} = 0$, and $\boldsymbol{K}_i \equiv 0$ when $i \notin \mathcal{U}$. Thus, the matrices $\bar{\boldsymbol{G}}$ and $\bar{\boldsymbol{K}}$ are structured, meaning that they have sparsity constraints determined by the topology of the SN, controller and actuators. Furthermore, since the matrix $\bar{\boldsymbol{L}}$ is block-diagonal, it can be viewed as a structured matrix with special sparsity constraint. In this sense, throughout this paper we will define $\mathcal{W}$ to be the set of all 3-tuples $(\bar{\boldsymbol{K}}, \bar{\boldsymbol{G}}, \bar{\boldsymbol{L}}) \in \mathbb{R}^{q_u\times np} \times \mathbb{R}^{np\times np} \times \mathbb{R}^{np\times\sum_{i\in\mathcal{S}}q_{y,i}}$ satisfying the sparsity constraints.

Let us choose a Lyapunov function candidate for the system (20) as

$$V(\tilde{\boldsymbol{x}}_s) = \tilde{\boldsymbol{x}}_s^T\boldsymbol{P}\tilde{\boldsymbol{x}}_s \tag{21}$$

where $\boldsymbol{P} > 0 \in \mathbb{R}^{n(p+1)\times n(p+1)}$. Calculating the time derivative of $V$ along the trajectory of the system (20), yields

$$\dot{V}(\tilde{\boldsymbol{x}}_s) = \tilde{\boldsymbol{x}}_s^T[\boldsymbol{P}\tilde{\boldsymbol{A}}_s + *]\tilde{\boldsymbol{x}}_s + 2\tilde{\boldsymbol{x}}_s^T\boldsymbol{P}(1_{p+1} \otimes \boldsymbol{I}_n)\boldsymbol{f}_s(\boldsymbol{x}_s, 0) + 2\tilde{\boldsymbol{x}}_s^T\boldsymbol{P}\tilde{\boldsymbol{B}}\tilde{\boldsymbol{w}} = \varsigma^T\boldsymbol{\Omega}_1\varsigma + \tau\|\boldsymbol{f}_s(\boldsymbol{x}_s, 0)\|^2 + \gamma^2\|\tilde{\boldsymbol{w}}\|^2 \tag{22}$$

where $\varsigma \triangleq \begin{bmatrix} \tilde{\boldsymbol{x}}_s \\ \boldsymbol{f}_s(\boldsymbol{x}_s, 0) \\ \tilde{\boldsymbol{w}} \end{bmatrix}$, $\boldsymbol{\Omega}_1 \triangleq \begin{bmatrix} [\boldsymbol{P}\tilde{\boldsymbol{A}}_s + *] & * & * \\ (1_{p+1} \otimes \boldsymbol{I}_n)^T\boldsymbol{P} & -\tau\boldsymbol{I} & * \\ \tilde{\boldsymbol{B}}^T\boldsymbol{P} & 0 & -\gamma^2\boldsymbol{I} \end{bmatrix}$

and $\tau > 0$ is a scalar. It is immediate from Assumption 2 that

$$\|\boldsymbol{f}_s(\boldsymbol{x}_s, 0)\| \le \kappa_1\|\boldsymbol{x}_s\| = \kappa_1\|\boldsymbol{H}_1\tilde{\boldsymbol{x}}_s\| \tag{23}$$

where $\boldsymbol{H}_1 = [\boldsymbol{I}_n \ \ 0 \ \cdots \ 0] \in \mathbb{R}^{n\times n(p+1)}$. Then, from (22) and (23), we have

$$\dot{V}(\tilde{\boldsymbol{x}}_s) \le \varsigma^T(\boldsymbol{\Omega}_1 + \boldsymbol{\Omega}_2)\varsigma + \gamma^2\|\tilde{\boldsymbol{w}}\|^2 \tag{24}$$

where $\boldsymbol{\Omega}_2 \triangleq \text{diag}\{\tau\kappa_1^2\boldsymbol{H}_1^T\boldsymbol{H}_1, 0, 0\}$.

Moreover, (17) can be rewritten as

$$\boldsymbol{u}(t) = \sum_{i\in\mathcal{S}}\boldsymbol{K}_i\hat{\boldsymbol{x}}_{s,i}(t) = \sum_{i\in\mathcal{S}}\boldsymbol{K}_i\boldsymbol{F}_i\tilde{\boldsymbol{x}}_s(t) = \bar{\boldsymbol{K}}\boldsymbol{F}\tilde{\boldsymbol{x}}_s(t) \tag{25}$$

where $\boldsymbol{F}_i = [\boldsymbol{I}_n \ \underbrace{0 \ \cdots \ 0}_{i-1} \ -\boldsymbol{I}_n \ \underbrace{0 \ \cdots \ 0}_{p-i}] \in \mathbb{R}^{n\times n(p+1)}$ and $\boldsymbol{F} = \text{col}_{i=1}^p\{\boldsymbol{F}_i\}$. Thus, from (24) and (25), it follows that

$$\dot{V}(\tilde{\boldsymbol{x}}_s) + \boldsymbol{x}_s^T\boldsymbol{Q}\boldsymbol{x}_s + \boldsymbol{u}^T\boldsymbol{R}\boldsymbol{u} - \gamma^2\tilde{\boldsymbol{w}}^T\tilde{\boldsymbol{w}} \le \varsigma^T\boldsymbol{\Lambda}\varsigma \tag{26}$$

where $\boldsymbol{\Lambda} = \boldsymbol{\Omega}_1 + \boldsymbol{\Omega}_2 + \boldsymbol{\Omega}_3$ and

$$\boldsymbol{\Omega}_3 \triangleq \text{diag}\{\boldsymbol{H}_1^T\boldsymbol{Q}\boldsymbol{H}_1 + (\boldsymbol{D}_R\bar{\boldsymbol{K}}\boldsymbol{F})^T\boldsymbol{D}_R\bar{\boldsymbol{K}}\boldsymbol{F}, 0, 0\}.$$

Obviously, if the following inequality holds:

$$\boldsymbol{\Lambda} < 0 \tag{27}$$

then we have

$$\dot{V}(\tilde{\boldsymbol{x}}_s) + \boldsymbol{x}_s^T\boldsymbol{Q}\boldsymbol{x}_s + \boldsymbol{u}^T\boldsymbol{R}\boldsymbol{u} - \gamma^2\tilde{\boldsymbol{w}}^T\tilde{\boldsymbol{w}} \le 0. \tag{28}$$

Therefore, we have the following result.

*Theorem 1:* Consider the system (20) where matrices $\bar{\boldsymbol{K}}$, $\bar{\boldsymbol{G}}$, and $\bar{\boldsymbol{L}}$ satisfy the given sparsity constraints, i.e., $(\bar{\boldsymbol{K}}, \bar{\boldsymbol{G}}, \bar{\boldsymbol{L}}) \in \mathcal{W}$. For some given $\gamma > 0$, if there exist a scalar $\tau > 0$ and a matrix $\boldsymbol{P} > 0$ satisfying (27), then the system (20) is exponentially stable in the absence of $\tilde{\boldsymbol{w}}$, and the $H_\infty$ control performance in (14) is guaranteed in the presence of $\tilde{\boldsymbol{w}}$ under zero-initial condition.

*Proof:* Assume that (27) holds for some $\tau > 0$ and $\boldsymbol{P} > 0$. Then, we have (28). Integrating (28) from $t = 0$ to $t = t_f$ yields

$$V(\tilde{\boldsymbol{x}}_s(t_f)) - V(\tilde{\boldsymbol{x}}_s(0)) + \int_0^{t_f} [\boldsymbol{x}_s^T(t)\boldsymbol{Q}\boldsymbol{x}_s(t) + \boldsymbol{u}^T(t)\boldsymbol{R}\boldsymbol{u}(t)]dt - \gamma^2 \int_0^{t_f} \tilde{\boldsymbol{w}}^T(t)\tilde{\boldsymbol{w}}(t)dt \le 0 . \quad (29)$$

Since $V(\tilde{\boldsymbol{x}}_s(0)) = 0$ under the zero-initial condition and $V(\tilde{\boldsymbol{x}}_s(t_f)) \ge 0$, we have (14) from (29).

Moreover, it is clear from (27) that

$$\begin{bmatrix} [\boldsymbol{P}\tilde{\boldsymbol{A}}_s + *] + \tau\kappa_1^2 \boldsymbol{H}_1^T \boldsymbol{H}_1 & * \\ (1_{p+1} \otimes \boldsymbol{I}_n)^T \boldsymbol{P} & -\tau \boldsymbol{I} \end{bmatrix} < 0$$

which implies that there exists a sufficiently small scalar $\sigma_1 > 0$ such that

$$\begin{bmatrix} [\boldsymbol{P}\tilde{\boldsymbol{A}}_s + *] + \tau\kappa_1^2 \boldsymbol{H}_1^T \boldsymbol{H}_1 & * \\ (1_{p+1} \otimes \boldsymbol{I}_n)^T \boldsymbol{P} & -\tau \boldsymbol{I} \end{bmatrix} \le -\sigma_1 \boldsymbol{I} . \quad (30)$$

Thus, by setting $\tilde{\boldsymbol{w}}(t) \equiv 0$, we have from (24) and (30) that

$$\dot{V}(\tilde{\boldsymbol{x}}_s) \le -\sigma_1[\tilde{\boldsymbol{x}}_s^T \tilde{\boldsymbol{x}}_s + \boldsymbol{f}_s^T(\boldsymbol{x}_s, 0)\boldsymbol{f}_s(\boldsymbol{x}_s, 0)]$$

which gives

$$\dot{V}(\tilde{\boldsymbol{x}}_s) \le -\sigma_1 \tilde{\boldsymbol{x}}_s^T \tilde{\boldsymbol{x}}_s \le -2\sigma_2 V(\tilde{\boldsymbol{x}}_s)$$

where $\sigma_2 = 0.5\sigma_1 / \lambda_{\max}(\boldsymbol{P})$. Thus, $V(\tilde{\boldsymbol{x}}_s(t)) \le V(\tilde{\boldsymbol{x}}_s(0))e^{-2\sigma_2 t}$, so that

$$\|\tilde{\boldsymbol{x}}_s(t)\| \le \sigma_3 e^{-\sigma_2 t} \|\tilde{\boldsymbol{x}}_s(0)\| \quad (31)$$

for all trajectories of $\tilde{\boldsymbol{x}}_s(t)$, where $\sigma_3 = \sqrt{\lambda_{\max}(\boldsymbol{P}) / \lambda_{\min}(\boldsymbol{P})}$. Hence, the system (20) with $\tilde{\boldsymbol{w}}(t) \equiv 0$ is exponentially stable. □

*Remark 4:* When the system (20) is exponentially stable in the absence of $\tilde{\boldsymbol{w}}$, it is clear that the estimation error dynamics in (19) is also stable, which means that the estimates of all local DCOs can converge to the actual state of the slow system exponentially.

Let us define $\rho \triangleq \gamma^2$ and partition $\boldsymbol{P}$ as

$$\boldsymbol{P} = \begin{bmatrix} \boldsymbol{P}_{00} & * & \cdots & * \\ \boldsymbol{P}_{10} & \boldsymbol{P}_{11} & \cdots & * \\ \vdots & \vdots & \ddots & \vdots \\ \boldsymbol{P}_{p0} & \boldsymbol{P}_{p1} & \cdots & \boldsymbol{P}_{pp} \end{bmatrix} > 0 \quad (32)$$

where $\boldsymbol{P}_{ij} \in \mathbb{R}^{n \times n}$, $i, j \in 0 \cup \mathcal{S}$, $j \le i$. Then, based on Theorem 1, we have the following theorem.

*Theorem 2:* Consider the PDE system (1)-(4). For some given scalar $\rho > 0$, suppose there exist a scalar $\tau > 0$, matrices $\boldsymbol{P}_{ij}$, $i, j \in 0 \cup \mathcal{S}$, $j \le i$, and matrices $\boldsymbol{K}_i$, $\boldsymbol{L}_i$, $\bar{\boldsymbol{G}}_{ij}$, $j \ne i \in \mathcal{S}$ satisfying LMI (32) and the BMI

$$\begin{bmatrix} \boldsymbol{\Xi}^{(1,1)} & * & * & * \\ \boldsymbol{\Xi}^{(2,1)} & -\tau \boldsymbol{I} & * & * \\ \boldsymbol{\Xi}^{(3,1)} & 0 & -\rho \boldsymbol{I} & * \\ \boldsymbol{\Xi}^{(4,1)} & 0 & 0 & -\boldsymbol{I} \end{bmatrix} < 0 \quad (33)$$

where

$$\boldsymbol{\Xi}^{(1,1)} \triangleq \begin{bmatrix} \boldsymbol{\Xi}_{00}^{(1,1)} + \tau\kappa_1^2 \boldsymbol{I}_n + \boldsymbol{Q} & * & * & * \\ \boldsymbol{\Xi}_{10}^{(1,1)} & \boldsymbol{\Xi}_{11}^{(1,1)} & * & * \\ \vdots & \vdots & \ddots & * \\ \boldsymbol{\Xi}_{p0}^{(1,1)} & \boldsymbol{\Xi}_{p1}^{(1,1)} & \cdots & \boldsymbol{\Xi}_{pp}^{(1,1)} \end{bmatrix},$$

$$\boldsymbol{\Xi}^{(2,1)} \triangleq \Big[ \sum_{i\in\{0\}} \boldsymbol{P}_{0i}^T + \sum_{i\in\{1,\cdots,p\}} \boldsymbol{P}_{i0} \quad \sum_{i\in\{0,1\}} \boldsymbol{P}_{1i}^T + \sum_{i\in\{2,\cdots,p\}} \boldsymbol{P}_{i1} \quad \cdots \quad \sum_{i\in\{0,\cdots,p-1\}} \boldsymbol{P}_{pi}^T + \sum_{i\in\{p\}} \boldsymbol{P}_{ip} \Big],$$

$$\boldsymbol{\Xi}^{(3,1)} \triangleq \begin{bmatrix} \boldsymbol{B}_{w,s}^T(\sum_{i\in\{0\}} \boldsymbol{P}_{0i}^T + \sum_{i\in\{1,\cdots,p\}} \boldsymbol{P}_{i0}) & -\boldsymbol{L}_1^T \boldsymbol{P}_{10} & \cdots & -\boldsymbol{L}_p^T \boldsymbol{P}_{p0} \\ \boldsymbol{B}_{w,s}^T(\sum_{i\in\{0,1\}} \boldsymbol{P}_{1i}^T + \sum_{i\in\{2,\cdots,p\}} \boldsymbol{P}_{i1}) & -\boldsymbol{L}_1^T \boldsymbol{P}_{11} & \cdots & -\boldsymbol{L}_p^T \boldsymbol{P}_{p1} \\ \vdots & \vdots & \ddots & \vdots \\ \boldsymbol{B}_{w,s}^T(\sum_{i\in\{0,\cdots,p-1\}} \boldsymbol{P}_{pi}^T + \sum_{i\in\{p\}} \boldsymbol{P}_{ip}) & -\boldsymbol{L}_1^T \boldsymbol{P}_{p1}^T & \cdots & -\boldsymbol{L}_p^T \boldsymbol{P}_{pp} \end{bmatrix}$$

$$\boldsymbol{\Xi}^{(4,1)} \triangleq \Big[ \boldsymbol{D}_R \sum_{i\in\mathcal{S}} \boldsymbol{K}_i \quad -\boldsymbol{D}_R \boldsymbol{K}_1 \quad \cdots \quad -\boldsymbol{D}_R \boldsymbol{K}_p \Big],$$

with

$$\boldsymbol{\Xi}_{00}^{(1,1)} = [\boldsymbol{P}_{00}\boldsymbol{A}_s + \boldsymbol{P}_{00}\boldsymbol{B}_{u,s} \sum_{l\in\mathcal{S}} \boldsymbol{K}_l + *],$$

$$\boldsymbol{\Xi}_{i0}^{(1,1)} = -\boldsymbol{K}_i^T \boldsymbol{B}_{u,s}^T \boldsymbol{P}_{00} + \boldsymbol{P}_{i0}\boldsymbol{B}_{u,s} \sum_{l\in\mathcal{S}} \boldsymbol{K}_l - \boldsymbol{C}_{s,i}^T \boldsymbol{L}_i^T \boldsymbol{P}_{i0} + \boldsymbol{P}_{i0}\boldsymbol{A}_s + \boldsymbol{A}_s^T \boldsymbol{P}_{i0}^T + \Big( \sum_{k\in\mathcal{S}} \bar{\boldsymbol{G}}_{ik}^T \boldsymbol{P}_{i0} - \sum_{k\in\mathcal{S}} \bar{\boldsymbol{G}}_{ki}^T \boldsymbol{P}_{k0} \Big),$$

$$\begin{aligned} \boldsymbol{\Xi}_{ij}^{(1,1)} = & -\boldsymbol{P}_{i0}\boldsymbol{B}_{u,s}\boldsymbol{K}_j - \boldsymbol{K}_i^T \boldsymbol{B}_{u,s}^T \boldsymbol{P}_{j0}^T + \boldsymbol{P}_{ij}\boldsymbol{A}_s + \boldsymbol{A}_s^T \boldsymbol{P}_{ij} \\ & -\boldsymbol{P}_{ij}\boldsymbol{L}_j\boldsymbol{C}_{s,j} - \boldsymbol{C}_{s,i}^T \boldsymbol{L}_i^T \boldsymbol{P}_{ij} + \Big( \boldsymbol{P}_{ij} \sum_{k\in\mathcal{S}} \bar{\boldsymbol{G}}_{jk} + \sum_{k\in\mathcal{S}} \bar{\boldsymbol{G}}_{ik}^T \boldsymbol{P}_{ij} \\ & - \sum_{l\in\{1,\cdots,i\}} \boldsymbol{P}_{il}\bar{\boldsymbol{G}}_{lj} - \sum_{l\in\{i+1,\cdots,p\}} \boldsymbol{P}_{li}^T \bar{\boldsymbol{G}}_{lj} \\ & - \sum_{l\in\{1,\cdots,j\}} \bar{\boldsymbol{G}}_{li}^T \boldsymbol{P}_{jl}^T - \sum_{l\in\{j+1,\cdots,p\}} \bar{\boldsymbol{G}}_{li}^T \boldsymbol{P}_{lj} \Big), \quad i, j \in \mathcal{S}, \ j \le i . \end{aligned}$$

Then there exists a controller of the form (13) based on the DCOs in (12) such that the augmented closed-loop system (20) is exponentially stable in the absence of $\tilde{\boldsymbol{w}}$, and the $H_\infty$ control performance in (14) is achieved with $\gamma = \sqrt{\rho}$ in the presence of disturbance $\tilde{\boldsymbol{w}}$. Furthermore, there exist positive real numbers $\eta_1^*$, $\eta_2^*$, $\varepsilon^*$ such that if $\|\boldsymbol{x}_s(0)\| \le \eta_1^*$, $\|\boldsymbol{x}_f(0)\|_{l^2} \le \eta_2^*$, $\varepsilon \in (0, \varepsilon^*)$, then the proposed controller guarantees that the

closed-loop PDE system is exponentially stable in the absence of disturbances $\boldsymbol{w}$ and $\boldsymbol{v}_i$, $i \in \mathcal{S}$. In this case, the consensus gains in (12) are given by

$$\boldsymbol{G}_{ij} = \frac{1}{m_{ij}} \bar{\boldsymbol{G}}_{ij}, \ i \in \mathcal{S}, \ j \in \mathcal{N}_i. \tag{34}$$

*Proof:* See Appendix A of [25]. □

*Remark 5:* Theorem 2 shows that the resulting finite dimensional DCOs-based $H_\infty$ control law can ensure that the closed-loop PDE system is exponentially stable in the absence of disturbances $\boldsymbol{w}$ and $\boldsymbol{v}_i$, $i \in \mathcal{S}$, provided that the initial condition and $\varepsilon \in (0, \varepsilon^*)$ are sufficiently small. This means that the spillover effect [7] can be tolerated by selecting a suitable eigenspectrum separation parameter $\varepsilon$.

*Remark 6:* It is observed that there exist many bilinear terms in (33) with respect to the decision variables $\boldsymbol{P}_{kl}$, $k, l \in 0 \cup \mathcal{S}$, $l \le k$, $\boldsymbol{K}_i$, $\boldsymbol{L}_i$, $\bar{\boldsymbol{G}}_{ij}$, $j \ne i \in \mathcal{S}$ such as $-\boldsymbol{L}_i^T \boldsymbol{P}_{jk}$ in $\boldsymbol{\Xi}^{(3,1)}$ and $\boldsymbol{P}_{00} \boldsymbol{B}_{u,s} \sum_{l \in \mathcal{S}} \boldsymbol{K}_l$ in $\boldsymbol{\Xi}_{00}^{(1,1)}$, and thus the inequality (33) is a BMI. Due to the non-convexity of BMIs, they are much more difficult to handle computationally than LMIs.

Based on Theorem 2, an optimal $H_\infty$ control design for the PDE system (1)-(4) can be formulated as the following optimization problem:

$$\min_{\mathcal{V}} \rho \ \text{ subject to matrix inequalities (32) and (33)} \tag{35}$$

where $\mathcal{V} \triangleq \{\tau, \rho, \boldsymbol{P}_{kl}, k, l \in 0 \cup \mathcal{S}, l \le k, \boldsymbol{K}_i, \boldsymbol{L}_i, \bar{\boldsymbol{G}}_{ij}, \ j \ne i \in \mathcal{S}\}$ is the set of decision variables. Obviously, the problem (35) is a BMI optimization problem, which is known to be NP-hard and cannot be solved efficiently by polynomial time interior-point methods [26]. So far, some local or global optimization approaches have been developed to solve the BMI problems, see, e.g., [27], [28] and references therein. In this paper, we solve the problem (35) using a local optimization algorithm as in [9] that treats the BMI as a double LMI, which can be directly solved by the existing LMI technique [29].

In order to find a feasible initial solution to start a local optimization for the problem (35), we subtract the matrix $\text{diag}\{\beta \boldsymbol{P}, 0, 0, 0, 0, 0\}$ with $\boldsymbol{P}$ given by (32), from the left of (33) for some given parameter $\beta > 0$ to obtain a necessary condition for the feasibility of (33). That is to say, if the inequality (33) holds for matrices $\boldsymbol{K}_i$, $\boldsymbol{L}_i$, $\bar{\boldsymbol{G}}_{ij}$, $j \ne i \in \mathcal{S}$, and $\boldsymbol{P} > 0$, then there exists a real number $\beta \ge 0$ such that the following inequality holds:

$$\begin{bmatrix} \boldsymbol{\Xi}^{(1,1)} - \beta \boldsymbol{P} & * & * & * \\ \boldsymbol{\Xi}^{(2,1)} & -\tau \boldsymbol{I} & * & * \\ \boldsymbol{\Xi}^{(3,1)} & 0 & -\rho \boldsymbol{I} & * \\ \boldsymbol{\Xi}^{(4,1)} & 0 & 0 & -\boldsymbol{I} \end{bmatrix} < 0 \tag{36}$$

Now, we will present the local optimization algorithm to give a suboptimal DCOs-based $H_\infty$ control design for PDE system (1)-(4) using an SN.

*Algorithm 1:*

*Step 1:* Choose sufficiently large scalars $\rho = \gamma^2 > 0$ and $\xi > 0$. Let $\boldsymbol{P}_{ii} = \xi \cdot \boldsymbol{I}$ for $i \in 0 \cup \mathcal{S}$ and $\boldsymbol{P}_{ij} = 0$ for $i, j \in 0 \cup \mathcal{S}$, $j < i$. Set $k = 0$, $l = 0$ and let $\rho_l = \rho$.

*Step 2:* Using $\boldsymbol{P}_{ij}$, $i, j \in 0 \cup \mathcal{S}$, $j \le i$, obtained in the previous step, solve the following LMI optimization problem for matrices $\boldsymbol{K}_i$, $\boldsymbol{L}_i$, $\bar{\boldsymbol{G}}_{ij}$, $j \ne i \in \mathcal{S}$, and scalars $\tau > 0$, $\beta$.

**OP 1:** Minimize $\beta$ subject to LMI (36).

If $\beta \le 0$, then go to Step 5. Otherwise, set $k = k + 1$ and go to Step 3.

*Step 3:* Using $\tau$, $\boldsymbol{K}_i$, $\boldsymbol{L}_i$ and $\bar{\boldsymbol{G}}_{ij}$, $j \ne i \in \mathcal{S}$ obtained in the previous step, solve the following LMI optimization problem for scalar $\beta$ and matrices $\boldsymbol{P}_{ij}$, $i, j \in 0 \cup \mathcal{S}$, $j \le i$:

**OP 2:** Minimize $\beta$ subject to LMIs (32) and (36).

If $\beta \le 0$, then go to Step 4. Otherwise, set $k = k + 1$ and go to Step 2.

*Step 4:* Using $\boldsymbol{P}_{ij}$, $i, j \in 0 \cup \mathcal{S}$, $j \le i$, obtained in previous step, solve the following LMI optimization problem for positive scalars $\rho$ and $\tau$, and matrices $\boldsymbol{K}_i$, $\boldsymbol{L}_i$, $\bar{\boldsymbol{G}}_{ij}$, $j \ne i \in \mathcal{S}$.

**OP 3:** Minimize $\rho$ subject to LMI (33).

Then set $l = l + 1$, $\rho_l = \rho$. If $|\rho_l - \rho_{l-1}| < \delta_\rho$, where $\delta_\rho$ is predetermined tolerance, go to Step 6; Else go to Step 5.

*Step 5:* Using $\tau$, $\boldsymbol{K}_i$, $\boldsymbol{L}_i$ and $\bar{\boldsymbol{G}}_{ij}$, $j \ne i \in \mathcal{S}$ obtained previously, solve the following LMI optimization problem for scalar $\rho > 0$ and matrices $\boldsymbol{P}_{ij}$, $i, j \in 0 \cup \mathcal{S}$, $j \le i$.

**OP 4:** Minimize $\rho$ subject to LMIs (32) and (33).

Then set $l = l + 1$, $\rho_l = \rho$. If $|\rho_l - \rho_{l-1}| < \delta_\rho$, go to Step 6; Else go to Step 4.

*Step 6:* A suboptimal solution of (35) is obtained and the optimized level is $\gamma_{opt} = \sqrt{\rho}$; STOP.

It is observed that Steps 1-3 of Algorithm 1 provide an iterative LMI algorithm to find an initially feasible solution for solving the BMI optimization problem (35) via (36). Clearly, when $\beta \le 0$ in Step 2 (or Step 3) of the algorithm, it implies that the resulting solution $\tau > 0$ and $\boldsymbol{K}_i$, $\boldsymbol{L}_i$ and $\bar{\boldsymbol{G}}_{ij}$, $j \ne i \in \mathcal{S}$ (or $\boldsymbol{P}_{ij}$, $i, j \in 0 \cup \mathcal{S}$, $j \le i$) also satisfies (32) and (33). Thus, a feasible initial solution to the problem (35) is obtained. As a consequence, Steps 4-6 of Algorithm 1 can be executed to find a suboptimal solution to the problem (35) in an iterative manner. It should be mentioned that one can change the parameter $\xi$ in Step 1 in order to obtain $\beta \le 0$ by Steps 1-3 of the algorithm. However, if $\beta \le 0$ cannot be obtained, Steps 1-3 fail to find a feasible initial solution to the problem (35). In this case, one must resort to other approaches.

## IV. Application to KSE System

In this section, we will consider the control problem of one dimensional KSE system with an SN to verify the effectiveness of the proposed method. The KSE system is described by the following nonlinear dissipative PDE:

$$\frac{\partial U(z,t)}{\partial t} = \mathcal{A}U(z,t) + f(U(z,t)) + \boldsymbol{b}_u^T(z)\boldsymbol{u}(t) + \boldsymbol{b}_w^T(z)\boldsymbol{w}(t) \quad (37)$$

subject to the periodic boundary conditions

$$\partial^j U(-\pi,t)/\partial z^j = \partial^j U(\pi,t)/\partial z^j, \ j = 0, 1, 2, 3 \quad (38)$$

and the initial condition

$$U(z,0) = 3\sin z + 2\sin 2z - \sin 3z \quad (39)$$

where $U(z,t)$ denotes the state variable, $z \in \Omega \triangleq [-\pi, \pi]$ is the spatial coordinate, $t$ is the time, $\mathcal{A} = -\vartheta \frac{\partial^4}{\partial z^4} - \frac{\partial^2}{\partial z^2}$ is a dissipative, linear spatial differential operator, $\vartheta$ is the instability parameter, and $f(U(z,t)) = -U(z,t)\frac{\partial U(z,t)}{\partial z}$ is the nonlinear function. $\boldsymbol{u}(t) \in \mathbb{R}^2$ is the manipulated input vector, $\boldsymbol{w}(t) \in \mathbb{R}^2$ is the process disturbance. The distribution functions $\boldsymbol{b}_u(z)$ and $\boldsymbol{b}_w(z)$ are respectively taken to be

$$\boldsymbol{b}_u(z) = \begin{bmatrix} \delta(z+0.2\pi) & \delta(z-0.4\pi) \end{bmatrix}^T,$$

$$\boldsymbol{b}_w(z) = \begin{bmatrix} \delta(z+0.1\pi) & \delta(z-0.2\pi) \end{bmatrix}^T.$$

The KSE system is measured via an SN with four nodes, whose measurement equations are given as

$$y_i(t) = \int_{-\pi}^{\pi} s_i(z)U(z,t)dz + v_i(t), \ i \in \mathcal{S} \triangleq \{1, 2, 3, 4\} \quad (40)$$

where $y_i(t) \in \mathbb{R}$ and $v_i(t) \in \mathbb{R}$ are the measured output and the measurement disturbance of the *i*-th node equipped with a single sensor. The distribution functions $s_i(z)$, $i \in \mathcal{S}$ are chosen as

$s_1(z) = \delta(z+0.6\pi)$, $s_2(z) = \delta(z+0.3\pi)$,

$s_3(z) = \delta(z-0.2\pi)$, $s_4(z) = \delta(z-0.5\pi)$.

These four nodes constitute an SN whose topology is represented by a directed graph $\mathcal{G} = (\mathcal{S}, \mathcal{E}, \mathcal{M})$ where $\mathcal{E}$ ={(1, 2), (1, 4), (2, 1), (3, 1), (4, 3)} and $\mathcal{M} = [m_{ij}]_{4\times 4}$ in which $m_{ij} = 1$ when $(i, j) \in \mathcal{E}$, and otherwise $m_{ij} = 0$.

The eigenvalue problem for the spatial differential operator of the KSE system of the form

$$\mathcal{A}U = -\vartheta \frac{\partial^4 U}{\partial z^4} - \frac{\partial^2 U}{\partial z^2},$$

$$\mathcal{D}(\mathcal{A}) \triangleq \{U \in \mathcal{H}_{2,[-\pi,\pi]} \text{ and } \frac{\partial^j U(-\pi)}{\partial z^j} = \frac{\partial^j U(\pi)}{\partial z^j}, \ j = 0, 1, 2, 3\}$$

can be solved analytically and its solution is given by

$$\lambda_j = -\vartheta j^4 + j^2, \ \phi_j(z) = \sin(jz)/\sqrt{\pi}, \ j = 1, 2, \cdots, \infty. \quad (41)$$

From (41), it can be found that when $\vartheta < 1$, there exist positive eigenvalues, i.e., the system (37) is unstable. Without loss of generality, we take $\vartheta = 0.4$ for the system (37) to show the effectiveness of the proposed method. For this system, we consider the first two eigenvalues as the dominant ones (and thus, $\varepsilon = |\lambda_1|/|\lambda_3| \approx 0.0256$ ). Then, a 2-dimensional slow system is derived as follows:

$$\dot{\boldsymbol{x}}_s(t) = \boldsymbol{A}_s\boldsymbol{x}_s(t) + \boldsymbol{B}_{u,s}\boldsymbol{u}(t) + \boldsymbol{f}_s(\boldsymbol{x}_s, 0) + \boldsymbol{B}_{w,s}\boldsymbol{w}(t) \quad (42)$$

with the measurement equations

$$y_i(t) = \boldsymbol{C}_{s,i}\boldsymbol{x}_s(t) + \bar{v}_i(t), \ i \in \mathcal{S} \quad (43)$$

where

$$\boldsymbol{x}_s = \begin{bmatrix} x_1 \\ x_2 \end{bmatrix}, \ \boldsymbol{x}_s(0) = \boldsymbol{x}_{s,0} = \begin{bmatrix} 5.3174 \\ 3.5449 \end{bmatrix}, \ \boldsymbol{A}_s = \mathrm{diag}\{0.6, -2.4\},$$

$$\boldsymbol{f}_s(\boldsymbol{x}_s, 0) = \begin{bmatrix} \langle f(\boldsymbol{\phi}_s^T(\cdot)\boldsymbol{x}_s), \phi_1(\cdot) \rangle \\ \langle f(\boldsymbol{\phi}_s^T(\cdot)\boldsymbol{x}_s), \phi_2(\cdot) \rangle \end{bmatrix}, \ \boldsymbol{\phi}_s(z) = \begin{bmatrix} \frac{1}{\sqrt{\pi}}\sin(z) \\ \frac{1}{\sqrt{\pi}}\sin(2z) \end{bmatrix},$$

$$\boldsymbol{B}_{u,s} = \begin{bmatrix} -0.3316 & 0.5366 \\ -0.5366 & 0.3316 \end{bmatrix}, \ \boldsymbol{B}_{w,s} = \begin{bmatrix} -0.1734 & 0.3316 \\ -0.3316 & 0.5366 \end{bmatrix},$$

$$\boldsymbol{C}_{s,1} = \begin{bmatrix} -0.5366 & 0.3316 \end{bmatrix}, \ \boldsymbol{C}_{s,2} = \begin{bmatrix} -0.4564 & -0.5366 \end{bmatrix},$$

$$\boldsymbol{C}_{s,3} = \begin{bmatrix} 0.3316 & 0.5366 \end{bmatrix}, \ \boldsymbol{C}_{s,4} = \begin{bmatrix} 0.5364 & 0 \end{bmatrix}.$$

Based on (42) and (43), the local DCOs of the SN are taken as

$$\dot{\hat{\boldsymbol{x}}}_{s,i} = \boldsymbol{A}_s\hat{\boldsymbol{x}}_{s,i} + \boldsymbol{B}_{u,s}\boldsymbol{u} + \boldsymbol{L}_i(y_i - \boldsymbol{C}_{s,i}\hat{\boldsymbol{x}}_{s,i}) + \sum_{i \in \mathcal{N}_i} \boldsymbol{G}_{ij}(\hat{\boldsymbol{x}}_{s,i} - \hat{\boldsymbol{x}}_{s,j}),$$

$$\hat{\boldsymbol{x}}_{s,i}(0) = 0, \ i \in \mathcal{S} \quad (44)$$

where $\mathcal{N}_1 = \{2, 4\}$, $\mathcal{N}_2 = \{1\}$, $\mathcal{N}_3 = \{1\}$, and $\mathcal{N}_4 = \{3\}$.

Assume that only the first node can transmit the state estimate of the slow system to controller, i.e., $\mathcal{U} = \{1\}$. Then we can adopt the following feedback control law:

$$\boldsymbol{u}(t) = \boldsymbol{K}_1\hat{\boldsymbol{x}}_{s,1}. \quad (45)$$

Let $\boldsymbol{Q} = \mathrm{diag}\{0.1, 0.1\}$ and $\boldsymbol{R} = 0$. Select $\rho = \xi = 900$, $\delta_\rho = 0.01$ in Algorithm 1. Running Steps 1-3 of the algorithm, we find that $\beta = -6.86$ for $k = 2$. Then continue the algorithm, i.e., run Steps 4-6 iteratively. When $l = 2$, the algorithm is terminated and a suboptimal solution of the optimization problem (35) is obtained as follows:

$$\tau = 88.8032, \ \rho_{\mathrm{DCO}} = 0.7980, \ \boldsymbol{K}_1 = \begin{bmatrix} -10.9647 & 5.9002 \\ -17.8487 & -0.9818 \end{bmatrix},$$

$$\boldsymbol{L}_1 = \begin{bmatrix} -15.8461 \\ 12.9702 \end{bmatrix}, \ \boldsymbol{L}_2 = \begin{bmatrix} -8.3935 \\ -12.5294 \end{bmatrix}, \ \boldsymbol{L}_3 = \begin{bmatrix} 6.7134 \\ 12.8918 \end{bmatrix},$$

$$\boldsymbol{L}_4 = \begin{bmatrix} 12.7132 \\ -4.2852 \end{bmatrix}, \ \boldsymbol{G}_{12} = \begin{bmatrix} 0.3431 & -7.7177 \\ -1.5052 & -8.5611 \end{bmatrix},$$

$$\boldsymbol{G}_{14} = \begin{bmatrix} -6.0126 & -1.5679 \\ 2.3492 & -4.1418 \end{bmatrix}, \ \boldsymbol{G}_{21} = \begin{bmatrix} -6.7274 & 5.5842 \\ 3.9281 & -4.3532 \end{bmatrix},$$

$$\boldsymbol{G}_{31} = \begin{bmatrix} -9.0659 & 3.2268 \\ 5.3902 & -4.4743 \end{bmatrix}, \ \boldsymbol{G}_{43} = \begin{bmatrix} -1.3052 & -2.6255 \\ 0.8020 & -7.9942 \end{bmatrix}.$$

Thus, we have $\gamma_{\text{DCO}} = \sqrt{\rho_{\text{DCO}}} = 0.8933$.

To compare with the proposed DCOs-based $H_\infty$ controller (*Controller 1*), a single observer (SO) based $H_\infty$ controller (*Controller 2*) is also considered, where the observer is chosen to be the first local one and the consensus gains $\boldsymbol{G}_{ij} \equiv 0$ in (44). Letting $i = j = 1$ and running Algorithm 1 yield the following solution for *Controller 2*:

$$\rho_{\text{SO}} = 4.9617\,,\ \boldsymbol{K}_1 = \begin{bmatrix} -6.5255 & 0.9326 \\ -10.8377 & 2.8401 \end{bmatrix},\ \boldsymbol{L}_1 = \begin{bmatrix} -11.7498 \\ 0.1848 \end{bmatrix},$$

$$\gamma_{\text{SO}} = \sqrt{\rho_{\text{SO}}} = 2.2275\,.$$

It is clear that $\gamma_{\text{DCO}} < \gamma_{\text{SO}}$, which implies that *Controller 1* can provide better $H_\infty$ control performance than *Controller 2*.

Now, we apply *Controllers 1* and *2* to the KSE system (37)-(39). Fig. 2 shows the closed-loop state evolution profiles of the disturbance-free KSE system under these two controllers, respectively. From Fig. 2 we observe that although both controllers can regulate the PDE state at the desired steady state $U(z,t) = 0$, *Controller 1* gives a faster convergence speed than *Controller 2*. Fig. 3 shows the actual state trajectory of the slow system and its estimates of the DCOs under *Controller 1*. Fig. 4 presents the actual state trajectory of the slow system and its estimate of the SO under *Controller 2*. It is observed from Figs. 3 and 4 that *Controller 1* can achieve faster state convergence of the slow system than *Controller 2*.

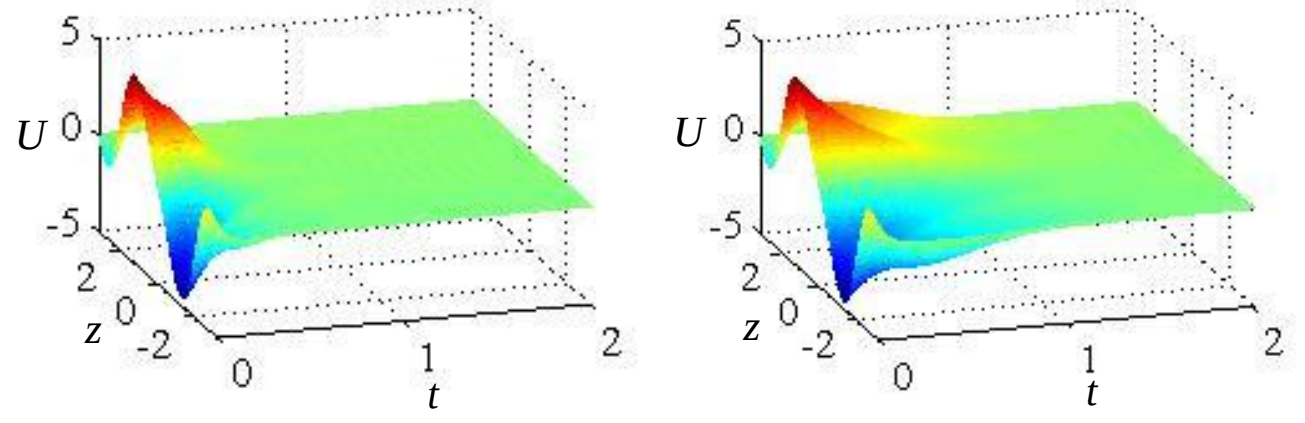

(a) *Controller 1* (b) *Controller 2*

Fig. 2 Closed-loop state evolution profiles of disturbance-free KSE system under two different controllers

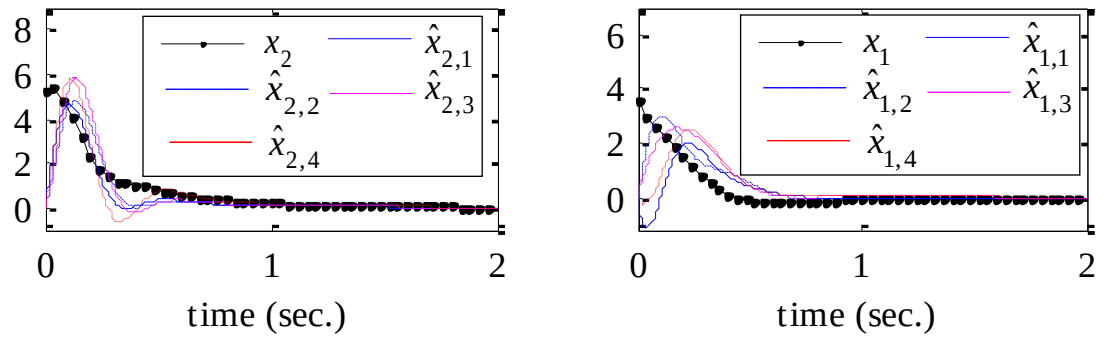

Fig. 3 Actual state trajectory of slow system and its estimates of DCOs under *Controller 1*

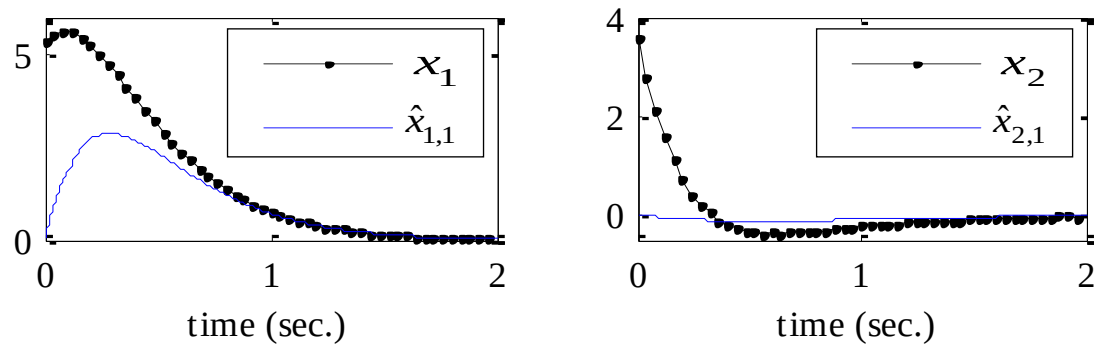

Fig. 4 Actual state trajectory of slow system and its estimate of SO under *Controller 2*

To verify the desired $H_\infty$ control performance, in the following simulation study, it is assumed that

$$\boldsymbol{w}(t) = \begin{bmatrix} 1.2\sin(20\pi t)e^{-0.01t} \\ -0.9\sin(40\pi t)e^{-0.02t} \end{bmatrix},\ v_1(t) = 0.6\cos(40\pi t)e^{-0.01t},$$

$$v_2(t) = -0.7\sin(60\pi t)e^{-0.02t},\ v_3(t) = 0.8\cos(80\pi t)e^{-0.03t},$$

$$v_4(t) = -0.9\sin(100\pi t)e^{-0.04t}.$$

Let $t_f = 4$ and define the following ratio:

$$J \triangleq \sqrt{\int_0^4 [\boldsymbol{x}_s^T(t)\boldsymbol{Q}\boldsymbol{x}_s(t) + \boldsymbol{u}^T(t)\boldsymbol{R}\boldsymbol{u}(t)]dt \Big/ \int_0^4 \tilde{\boldsymbol{w}}^T(t)\tilde{\boldsymbol{w}}(t)dt}$$

where $\tilde{\boldsymbol{w}} = [\boldsymbol{w}^T\ \bar{\boldsymbol{v}}^T]^T$ with $\bar{\boldsymbol{v}} = [\bar{v}_1\ \cdots\ \bar{v}_4]^T$ for *Controller 1* and $\bar{\boldsymbol{v}} = \bar{v}_1$ for *Controller 2*, and $\bar{v}_i(t) = y_i(t) - \boldsymbol{C}_{s,i}\boldsymbol{x}_s(t)$ for $i \in \mathcal{S}$. Then, we can obtain that $J_{\text{DCO}} = 0.0027 < \gamma_{\text{DCO}} = 0.8933$ for *Controller 1* and $J_{\text{SO}} = 0.0040 < \gamma_{\text{SO}} = 2.2275$ for *Controller 2*. Moreover, it is observed that $J_{\text{DCO}} < J_{\text{SO}}$, which implies that *Controller 1* has a stronger ability of disturbance attenuation than *Controller 2*.

## V. Conclusions

In this paper, the finite dimensional DCOs-based $H_\infty$ control problem has been addressed for a class of nonlinear dissipative PDE systems with SNs of given topology. The modal decomposition and singular perturbation techniques are initially applied to the PDE system to derive a slow system of finite dimensional ODEs. Then, based on the slow system, a set of finite dimensional DCOs are constructed to implement a centralized control scheme which only uses the available estimates from the specified group of SN nodes. A BMI-based $H_\infty$ control design method is developed such that the original closed-loop PDE system is exponentially stable and a prescribed level of disturbance attenuation is satisfied for the slow system. Moreover, by treating the BMI as double LMI, a local optimization algorithm is proposed to give a suboptimal $H_\infty$ controller such that the attenuation level is made as small as possible. Finally, the simulation results on the control of one dimensional KSE system indicate that the proposed design method is effective.

## Acknowledgment

The authors gratefully acknowledge the helpful comments and suggestions of the Associate Editor and anonymous reviewers, which have improved the presentation of this paper.

## Appendix A

*Proof of Theorem 2:* By considering (32) and matrices $\tilde{\boldsymbol{A}}$ and $\tilde{\boldsymbol{B}}$ in (20), $\boldsymbol{\Xi}^{(1,1)}$, $\boldsymbol{\Xi}^{(2,1)}$, $\boldsymbol{\Xi}^{(3,1)}$, and $\boldsymbol{\Xi}^{(4,1)}$ in (33) can be respectively rewritten as

$$\boldsymbol{\Xi}^{(1,1)} = [\boldsymbol{P}\tilde{\boldsymbol{A}}_s + *] + \tau\kappa_1^2\boldsymbol{H}_1^T\boldsymbol{H}_1 + \boldsymbol{H}_1^T\boldsymbol{Q}\boldsymbol{H}_1\,,$$

$$\boldsymbol{\Xi}^{(2,1)} = (1_{p+1} \otimes \boldsymbol{I}_n)^T\boldsymbol{P}\,,\ \boldsymbol{\Xi}^{(3,1)} = \tilde{\boldsymbol{B}}^T\boldsymbol{P}\,,\ \boldsymbol{\Xi}^{(4,1)} = \boldsymbol{D}_R\bar{\boldsymbol{K}}\boldsymbol{F}$$

which mean that (33) can be written as

$$\begin{bmatrix} [\boldsymbol{P}\tilde{\boldsymbol{A}}_s + *] + \tau\kappa_1^2 \boldsymbol{H}_1^T \boldsymbol{H}_1 + \boldsymbol{H}_1^T \boldsymbol{Q}\boldsymbol{H}_1 & * & * & * \\ (1_{p+1} \otimes \boldsymbol{I}_n)^T \boldsymbol{P} & -\tau \boldsymbol{I} & * & * \\ \tilde{\boldsymbol{B}}^T \boldsymbol{P} & 0 & -\rho \boldsymbol{I} & * \\ \boldsymbol{D}_R \bar{\boldsymbol{K}}\boldsymbol{F} & 0 & 0 & -\boldsymbol{I} \end{bmatrix} < 0 . \quad \text{(A1)}$$

By Schur complement, it follows that (A1) is equivalent to (27). This implies that for the 3-tuples of matrices $(\bar{\boldsymbol{K}}, \bar{\boldsymbol{G}}, \bar{\boldsymbol{L}}) \in \mathcal{W}$ consisting of $\boldsymbol{K}_i$, $\boldsymbol{L}_i$, $\bar{\boldsymbol{G}}_{ij}$, $j \neq i \in \mathcal{S}$, there exist the scalar $\tau > 0$ and the matrix $\boldsymbol{P} > 0$ consisting of $\boldsymbol{P}_{ij}$, $i, j \in 0 \cup \mathcal{S}$, $j \le i$, satisfying (27). Thus, we can conclude from Theorem 1 that the closed-loop system (20) is exponentially stable in the absence of disturbance $\tilde{\boldsymbol{w}}$, and the $H_\infty$ control performance (14) with $\gamma = \sqrt{\rho}$ is satisfied in the presence of $\tilde{\boldsymbol{w}}$.

Next, we will show that the closed-loop PDE system is exponentially stable when $\boldsymbol{w}(t) \equiv 0$ and $\boldsymbol{v}_i \equiv 0$, $i \in \mathcal{S}$, provided that the initial condition and $\varepsilon \triangleq |\lambda_L| / |\lambda_{n+1}|$ are sufficiently small. Setting $\boldsymbol{w}(t) \equiv 0$ and $\boldsymbol{v}_i \equiv 0$, $i \in \mathcal{S}$ in (7) and (19), substituting (13) into (7) and considering (19), yield the following augmented system:

$$\begin{cases} \dot{\boldsymbol{x}}_s = \boldsymbol{A}_s \boldsymbol{x}_s + \boldsymbol{B}_{u,s} \sum_{i \in \mathcal{S}} \boldsymbol{K}_i \hat{\boldsymbol{x}}_{s,i} + \boldsymbol{f}_s(\boldsymbol{x}_s, \boldsymbol{x}_f) \\ \dot{\bar{\boldsymbol{e}}}_s = (\bar{\boldsymbol{A}}_s - \bar{\boldsymbol{L}}\bar{\boldsymbol{C}}_s + \bar{\boldsymbol{G}})\bar{\boldsymbol{e}}_s + (1_p \otimes \boldsymbol{I}_n) \boldsymbol{f}_s(\boldsymbol{x}_s, 0) - \bar{\boldsymbol{L}}\bar{\boldsymbol{y}}_f \\ \dot{\boldsymbol{x}}_f = \boldsymbol{A}_f \boldsymbol{x}_f + \boldsymbol{B}_{u,f} \sum_{i \in \mathcal{S}} \boldsymbol{K}_i \hat{\boldsymbol{x}}_{s,i} + \boldsymbol{f}_f(\boldsymbol{x}_s, \boldsymbol{x}_f) \end{cases} \quad \text{(A2)}$$

where $\bar{\boldsymbol{y}}_f \triangleq \begin{bmatrix} \boldsymbol{y}_{f,1}^T & \cdots & \boldsymbol{y}_{f,p}^T \end{bmatrix}^T \in \mathbb{R}^{\sum_{i \in \mathcal{S}} q_{y,i}}$, which can be rewritten as

$$\begin{cases} \dot{\tilde{\boldsymbol{x}}}_s = \tilde{\boldsymbol{A}}\tilde{\boldsymbol{x}}_s + (1_{p+1} \otimes \boldsymbol{I}_n) \boldsymbol{f}_s(\boldsymbol{x}_s, 0) + \begin{bmatrix} \boldsymbol{f}_s(\boldsymbol{x}_s, \boldsymbol{x}_f) - \boldsymbol{f}_s(\boldsymbol{x}_s, 0) \\ -\bar{\boldsymbol{L}}\bar{\boldsymbol{y}}_f \end{bmatrix} \\ \dot{\boldsymbol{x}}_f = \boldsymbol{A}_f \boldsymbol{x}_f + \boldsymbol{B}_{u,f} \bar{\boldsymbol{K}}\boldsymbol{F}\tilde{\boldsymbol{x}}_s + \boldsymbol{f}_f(\boldsymbol{x}_s, \boldsymbol{x}_f) \end{cases} \quad \text{(A3)}$$

where $\tilde{\boldsymbol{x}}_s$ is defined in (20). Noting the condition that $f(\bar{x}(z,t))$ is locally Lipschitz continuous, we have that $\boldsymbol{f}_s(\boldsymbol{x}_s, \boldsymbol{x}_f)$ and $\boldsymbol{f}_f(\boldsymbol{x}_s, \boldsymbol{x}_f)$ are also Lipschitz continuous. Thus, for some given positive real numbers $\eta_1^*$, $\eta_2^*$ such that $\|\boldsymbol{x}_s\| \le \eta_1^*$ and $\|\boldsymbol{x}_f\|_{l^2} \le \eta_2^*$, then there exist positive real numbers $\kappa_2$, $\kappa_3$, $\kappa_4$ such that

$$\begin{cases} \|\boldsymbol{f}_s(\boldsymbol{x}_s, \boldsymbol{x}_f) - \boldsymbol{f}_s(\boldsymbol{x}_s, 0)\| \le \kappa_2 \|\boldsymbol{x}_f\|_{l^2} \\ \|\boldsymbol{f}_f(\boldsymbol{x}_s, \boldsymbol{x}_f)\|_{l^2} \le \kappa_3 \|\boldsymbol{x}_s\| + \kappa_4 \|\boldsymbol{x}_f\|_{l^2} \end{cases} \quad \text{(A4)}$$

Pick $a_4 < \eta_1^*$ and $b_4 < \eta_2^*$. Since the closed-loop system (20) is exponentially stable in the absence of $\tilde{\boldsymbol{w}}$, from the converse Lyapunov theorem, we have that there exists a smooth Lyapunov function $V_s : \mathbb{R}^{n(p+1)} \to \mathbb{R}_+$ and a set of numbers $a_1$, $a_2$, $a_3$, $a_4$, $a_5$ such that for all $\tilde{\boldsymbol{x}}_s \in \mathbb{R}^{n(p+1)}$ satisfying $\|\tilde{\boldsymbol{x}}_s\| \le a_4$, the following conditions hold:

$$\begin{cases} a_1 \|\tilde{\boldsymbol{x}}_s\|^2 \le V_s(\tilde{\boldsymbol{x}}_s) \le a_2 \|\tilde{\boldsymbol{x}}_s\|^2 \\ \dot{V}_s(\tilde{\boldsymbol{x}}_s) = \dfrac{\partial V_s(\tilde{\boldsymbol{x}}_s)}{\partial \tilde{\boldsymbol{x}}_s} [\tilde{\boldsymbol{A}}\tilde{\boldsymbol{x}}_s + (1_{p+1} \otimes \boldsymbol{I}_n) \boldsymbol{f}_s(\boldsymbol{x}_s, 0)] \le -a_3 \|\tilde{\boldsymbol{x}}_s\|^2 \\ \left\| \dfrac{\partial V_s(\tilde{\boldsymbol{x}}_s)}{\partial \tilde{\boldsymbol{x}}_s} \right\| \le a_5 \|\tilde{\boldsymbol{x}}_s\| \end{cases} \quad \text{(A5)}$$

By considering the orthogonality of eigenfunctions and $k_u \boldsymbol{b}_u(z) = \boldsymbol{\phi}_s^T(z) \boldsymbol{B}_{u,s} + \boldsymbol{\phi}_f^T(z) \boldsymbol{B}_{u,f}$, it follows that

$$\int_\Omega \|k_u \boldsymbol{b}_u(z)\|^2 dz = \boldsymbol{B}_{u,s}^T \boldsymbol{B}_{u,s} + \boldsymbol{B}_{u,f}^T \boldsymbol{B}_{u,f} .$$

Thus, we have

$$\upsilon_1 \triangleq \bar{\sigma}(\boldsymbol{B}_{u,f}) = \lambda_{\max}^{\frac{1}{2}}(\boldsymbol{B}_{u,f}^T \boldsymbol{B}_{u,f}) = \lambda_{\max}^{\frac{1}{2}}\left( \int_\Omega \|k_u \boldsymbol{b}_u(z)\|^2 dz - \boldsymbol{B}_{u,s}^T \boldsymbol{B}_{u,s} \right) \quad \text{(A6)}$$

Let us define the following induced norms:

$$\varpi_i \triangleq \sup_{0 \neq \boldsymbol{x}_f \in l^2} \frac{\|\boldsymbol{y}_{f,i}\|}{\|\boldsymbol{x}_f\|_{l^2}} = \sup_{(\|\bar{x}(\cdot,t)\|_{2,\Omega}^2 - \|x_s(t)\|^2) \neq 0} \frac{\left\| \int_\Omega \boldsymbol{S}_i(z) \bar{\boldsymbol{x}}(z,t) dz - \boldsymbol{C}_{s,i} \boldsymbol{x}_s(t) \right\|}{\left( \|\bar{\boldsymbol{x}}(\cdot,t)\|_{2,\Omega}^2 - \|\boldsymbol{x}_s(t)\|^2 \right)^{\frac{1}{2}}} \ge 0, \ i \in \mathcal{S}$$

where the fact $\|\bar{\boldsymbol{x}}(\cdot,t)\|_2^2 = \|\boldsymbol{x}_s(t)\|^2 + \|\boldsymbol{x}_f(t)\|_{l^2}^2$ has been used. Thus, $\|\boldsymbol{y}_{f,i}\| \le \varpi_i \|\boldsymbol{x}_f\|_{l^2}$, $i \in \mathcal{S}$, which imply that

$$\|\bar{\boldsymbol{y}}_f\| = \sqrt{\sum_{i=1}^p \|\bar{\boldsymbol{y}}_{f,i}\|^2} \le \bar{\varpi} \|\boldsymbol{x}_f\|_{l^2} \quad \text{(A7)}$$

where $\bar{\varpi} \triangleq \sqrt{\sum_{i=1}^p \varpi_i^2}$.

Consider the smooth function $V : \mathbb{R}^{n(p+1)} \times l^2 \to \mathbb{R}_+$ given by

$$V(\tilde{\boldsymbol{x}}_s, \boldsymbol{x}_f) = V_s(\tilde{\boldsymbol{x}}_s) + 0.5 q_f \boldsymbol{x}_f^T \boldsymbol{x}_f \quad \text{(A8)}$$

as a Lyapunov function candidate for system (A3) where $q_f > 0$ is some given constant. Computing the time derivative of $V(\tilde{\boldsymbol{x}}_s, \boldsymbol{x}_f)$ along the trajectories of system (A3), and considering (A4)-(A7), give

$$\dot{V}(\tilde{\boldsymbol{x}}_s, \boldsymbol{x}_f) = \frac{\partial V_s(\tilde{\boldsymbol{x}}_s)}{\partial \tilde{\boldsymbol{x}}_s} \dot{\tilde{\boldsymbol{x}}}_s + q_f \boldsymbol{x}_f^T \dot{\boldsymbol{x}}_f$$

$$\le \frac{\partial V_s(\tilde{\boldsymbol{x}}_s)}{\partial \tilde{\boldsymbol{x}}_s} [\tilde{\boldsymbol{A}}\tilde{\boldsymbol{x}}_s + (1_{p+1} \otimes \boldsymbol{I}_n) \boldsymbol{f}_s(\boldsymbol{x}_s, 0)]$$

$$+ \frac{\partial V_s(\tilde{\boldsymbol{x}}_s)}{\partial \tilde{\boldsymbol{x}}_s} \begin{bmatrix} \boldsymbol{f}_s(\boldsymbol{x}_s, \boldsymbol{x}_f) - \boldsymbol{f}_s(\boldsymbol{x}_s, 0) \\ -\bar{\boldsymbol{L}}\bar{\boldsymbol{y}}_f \end{bmatrix} + q_f \boldsymbol{x}_f^T \boldsymbol{A}_f \boldsymbol{x}_f$$

$$+ q_f \boldsymbol{x}_f^T [\boldsymbol{f}_f(\boldsymbol{x}_s, \boldsymbol{x}_f) + \boldsymbol{B}_{u,f} \bar{\boldsymbol{K}}\boldsymbol{F}\tilde{\boldsymbol{x}}_s]$$

$$\leq -a_3\|\tilde{\boldsymbol{x}}_s\|^2 + a_5\|\tilde{\boldsymbol{x}}_s\|\left(\kappa_2\|\boldsymbol{x}_f\|_{l^2} + \upsilon_2\varpi\|\boldsymbol{x}_f\|_{l^2}\right) + q_f\lambda_{n+1}\|\boldsymbol{x}_f\|_{l^2}^2$$
$$+q_f\|\boldsymbol{x}_f\|_{l^2}\left(\kappa_3\|\boldsymbol{x}_s\| + \kappa_4\|\boldsymbol{x}_f\|_{l^2} + \upsilon_1\upsilon_3\|\tilde{\boldsymbol{x}}_s\|\right)$$
$$\leq -a_3\|\tilde{\boldsymbol{x}}_s\|^2 + \upsilon_4\|\tilde{\boldsymbol{x}}_s\|\|\boldsymbol{x}_f\|_{\mathbb{R}^\infty} + q_f(\lambda_{n+1}+\kappa_4)\|\boldsymbol{x}_f\|_{l^2}^2$$
$$= -\begin{bmatrix}\|\tilde{\boldsymbol{x}}_s\| & \|\boldsymbol{x}_f\|_{l^2}\end{bmatrix}\Upsilon\begin{bmatrix}\|\tilde{\boldsymbol{x}}_s\| \\ \|\boldsymbol{x}_f\|_{l^2}\end{bmatrix}$$

where $\upsilon_2 \triangleq \lambda_{\max}^{\frac{1}{2}}(\bar{\boldsymbol{L}}^T\bar{\boldsymbol{L}})$, $\upsilon_3 \triangleq \lambda_{\max}^{\frac{1}{2}}(\boldsymbol{F}^T\bar{\boldsymbol{K}}^T\bar{\boldsymbol{K}}\boldsymbol{F})$,
$\upsilon_4 \triangleq a_5(\kappa_2+\upsilon_2\varpi)+q_f(\kappa_3+\upsilon_1\upsilon_3)$,
$$\Upsilon = \begin{bmatrix} a_3 & * \\ -0.5\upsilon_4 & -q_f(\lambda_{n+1}+\kappa_4)\end{bmatrix}.$$

By considering the fact $\lambda_{n+1} = -\varepsilon^{-1}|\lambda_L|$ and defining

$$\varepsilon^* \triangleq \frac{|\lambda_L| a_3 q_f}{a_3 q_f \kappa_4 + 0.25\upsilon_4^2}, \tag{A9}$$

we have that if $\varepsilon \in (0,\varepsilon^*)$ , then $\Upsilon > 0$ and thus $\dot{V}(\tilde{\boldsymbol{x}}_s,\boldsymbol{x}_f) \leq -\lambda_{\min}(\Upsilon)(\|\tilde{\boldsymbol{x}}_s\|^2 + \|\boldsymbol{x}_f\|_{l^2}^2)$ , which directly implies that the system (A3) is exponentially stable. Obviously, this implies that the system (7) is also exponentially stable. Then, the exponential stability of the system (7) implies that the closed-loop PDE system is exponentially stable. For example, for $\bar{x}(z,t) \in \mathbb{R}$ , if $\|\boldsymbol{x}(t)\|_{l^2} \leq c_2 e^{-c_3 t}\|\boldsymbol{x}(0)\|_{l^2}$ , $\forall t \geq 0$ for all $\boldsymbol{x}(t) = [\boldsymbol{x}_s^T(t) \quad \boldsymbol{x}_f^T(t)]^T \in l^2$ satisfying $\|\boldsymbol{x}(t)\|_{l^2} \leq c_1$ , where $c_1$ , $c_2$ and $c_3$ are positive real numbers, then considering the fact $\|\bar{x}(\cdot,t)\|_{2,\Omega} = \|\boldsymbol{x}(t)\|_{l^2}$ yields $\|\bar{x}(\cdot,t)\|_{2,\Omega} \leq c_2 e^{-c_3 t}\|\bar{x}_0(\cdot)\|_{2,\Omega}$ , $\forall t \geq 0$ for all $\bar{x}(z,t) \in \mathcal{H}_{2,\Omega}$ satisfying $\|\bar{x}(\cdot,t)\|_{2,\Omega} \leq c_1$ . This completes the proof. □

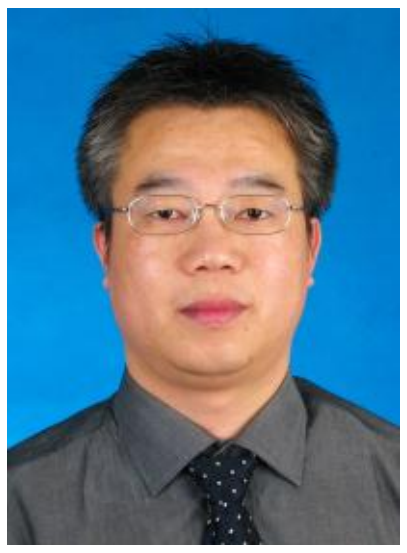

**Huai-Ning Wu** was born in Anhui, China, on November 15, 1972. He received the B.E. degree in automation from Shandong Institute of Building Materials Industry, Jinan, China and the Ph.D. degree in control theory and control engineering from Xi'an Jiaotong University, Xi'an, China, in 1992 and 1997, respectively.

From August 1997 to July 1999, he was a Postdoctoral Researcher in the Department of Electronic Engineering at Beijing Institute of Technology, Beijing, China. In August 1999, he joined the School of Automation Science and Electrical Engineering, Beihang University (formerly Beijing University of Aeronautics and Astronautics), Beijing. From December 2005 to May 2006, he was a Senior Research Associate with the Department of Manufacturing Engineering and Engineering Management (MEEM), City University of Hong Kong, Kowloon, Hong Kong. From October to December during 2006-2008 and from July to August in 2010, he was a Research Fellow with the Department of MEEM, City University of Hong Kong. From July to August in 2011 and 2013, he was a Research Fellow with the Department of Systems Engineering and Engineering Management, City University of Hong Kong. He is currently a Professor with Beihang University. His current research interests include robust control, fault-tolerant control, distributed parameter systems, and fuzzy/neural modeling and control.

Dr. Wu serves as Associate Editor of the *IEEE Transactions on Systems, Man & Cybernetics: Systems*. He is a member of the Committee of Technical Process Failure Diagnosis and Safety, Chinese Association of Automation.

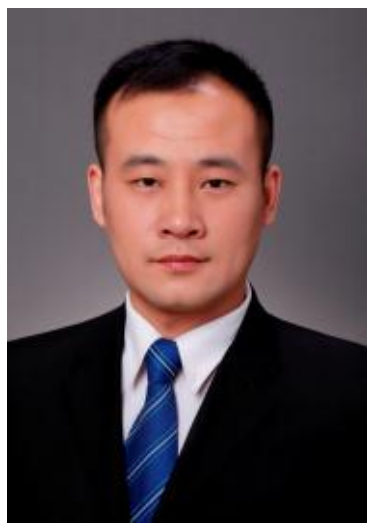

**Hong-Du Wang** was born in Shandong Province, China, on May 5, 1984. He received the B.S. degree in automation and the M.S. degree in Control theory and Control Engineering both from Ocean University of China (OUC), Qingdao, China, in 2006 and 2011, respectively. He is studying for the Ph.D. degree in Control Science and Engineering from Beihang University (Beijing University of Aeronautics and Astronautics) in China.

His current research interests include distributed parameter systems, sensor networks and anti-disturbance control.